\documentclass[aps,notitlepage,superscriptaddress,floatfix,pra,twocolumn]{revtex4-1}

\usepackage{lipsum}
\usepackage{graphicx}
\usepackage{amsmath,amssymb}
\usepackage{braket}
\usepackage{mathtools}
\usepackage{hyperref}
\usepackage{multirow}
\usepackage{color}
\usepackage[normalem]{ulem}
\usepackage{amsfonts}
\usepackage{float}
\usepackage{pdfpages} 
\makeatletter
\usepackage{subfigure}
\usepackage{soul, xcolor}
\usepackage{dsfont}

\setstcolor{red}

\newcommand{\cP}{\mathcal P}
\newcommand{\cA}{\mathcal A}

\newcommand{\dsym}{\boldsymbol{\mathrm d}}
\newcommand{\usym}{\boldsymbol{\mathrm u}}
\newcommand{\rsym}{\boldsymbol{\mathrm r}}

\newcommand{\tvoid}{\tau_{\rm void}}

\newcommand{\Ord}{\mathcal O}

\def\bra#1{\mathinner{\langle{#1}|}}
\def\ket#1{\mathinner{|{#1}\rangle}}

\def\beq{\begin{equation}}
\def\eeq{\end{equation}}
\def\bea{\begin{eqnarray}}
\def\eea{\end{eqnarray}}

\usepackage{pdfpages} 
\makeatletter
\AtBeginDocument{\let\LS@rot\@undefined}
\makeatother

\begin{document}

\title{KPZ Superdiffusion of Local Correlators in Diffusive Random Quantum Circuits}

\author{Ewan McCulloch}
\affiliation{Laboratoire de Physique de l'École Normale Supérieure, CNRS, ENS \& Université PSL; 24 rue Lhomond, 75005 Paris, France}

\begin{abstract}
We study the single-particle Green's function \(G(x,t)=\langle \sigma^-_x(0)\sigma^+_0(t)\rangle\) in one-dimensional particle-number-conserving random unitary circuits coupled to an external bath. For fixed spacetime disorder, we argue that \(G(x,t)\) is governed, in both the strong- and weak-noise limits, by directed waves in a random medium. We find Kardar--Parisi--Zhang (KPZ) scaling in the wandering statistics of the normalized spatial distribution \(p(x,t)\propto |G(x,t)|^2\) and in the associated free energy. In particular, its center \(\langle x(t)\rangle\equiv\sum_x x\,p(x,t)\) wanders on a length-scale \(\Ord(t^{2/3})\), while sample-to-sample fluctuations of \(-\log\sum_x |G(x,t)|^2\) scale as \(t^{1/3}\). At weak noise \(\gamma\ll1\), the crossover to the strong-disorder fixed point occurs at a parametrically long time \(\Ord(\gamma^{-3/2})\). These predictions are confirmed numerically using tensor-network simulations of the noisy operator dynamics in individual circuits at moderate noise, and of a phase-annealed proxy retaining hopping disorder at weak noise.
\end{abstract}

\maketitle

\textbf{\textit{Introduction}}---Understanding how isolated quantum many-body systems thermalize remains amcentral problem in nonequilibrium physics. The standard assumption is that local degrees of freedom rapidly equilibrate, subject only to local conservation laws, while the subsequent dynamics of the conserved densities is described by hydrodynamics.  This picture has been remarkably successful~\cite{PhysRevA.89.053608, wienand2023emergence, joshi2022observing, rosenberg2023dynamics, wei2022quantum, le2023observation, gross2017quantum, scholl2021microwave, heavyionlecturenotes,heavyionreview, Muller2008,Lucas2016a,Lucas2016b,Crossno2016,Narozhny2017,Lucas2018,Bal2021,McCulloch2023,SinghNavierStokes,gopalakrishnan2024non,2026arXiv260102475M,Bauer2017,Bernard_2019,QSSEP_Bernard_2021,2026arXiv260116883A}, but hydrodynamics is a theory of effectively classical variables such as densities and currents, not of intrinsically quantum, phase-sensitive observables.

Single-particle Green's functions are a basic example. They underlie experimental probes such as photoemission spectroscopy, scanning tunneling microscopy, and spin-echo NMR~\cite{RevModPhys.93.025006,2010Sci...329.1628L,2023AAOM....1..924L,PhysRevB.83.064302,PhysRevB.86.214410}, and more generally diagnose the loss of locally recoverable quantum information in chaotic dynamics. (By contrast, out-of-time-ordered correlators and related operator-spreading diagnostics are nonlocal probes of information spreading~\cite{Maldacena2016,PhysRevLett.117.091602,Nahum_18,von_Keyserlingk_2018,Rakovszky2018,Khemani_2018}.) A standard intuition is that a chaotic many-body system acts as its own bath~\cite{BAA_06}: since phase-sensitive correlators such as Green's functions decay exponentially in the presence of an external bath~\cite{StretchedExp}, one might expect the same exponential decay in isolated chaotic systems.

This expectation was recently shown to fail. Quantum coherent motion persists longer than expected, supported in rare low-entropy regions, yielding a stretched-exponential rather than exponential decay of Green's functions $G(x,t)\equiv\langle \sigma^-_x(0)\sigma^+_0(t)\rangle$ ~\cite{StretchedExp,previous}. 
Those results, however, were largely annealed: for example, studying ensemble averages \(\mathbb{E}[|G|^2]\) in random circuit models. The universal long-time behavior of Green's functions in typical circuit realizations has yet to be determined.

We study $U(1)$-conserving random unitary circuits in one dimension coupled to an external bath, and compute the Green's function in fixed realizations.  We find that \(G\) is governed by a strong-disorder fixed point. In the strong-noise limit we exploit an exact mapping to directed-waves in a random medium~\cite{PhysRevLett.62.941,PhysRevLett.66.2176,PhysRevA.45.8859,BLUM1992588,1994cond.mat.11022K,PhysRevB.83.052405} (DWRM), predicting Kardar-Parisi-Zhang (KPZ) fluctuations~\cite{KPZ,HALPINHEALY1995215}: sample-to-sample spatial wandering scales as \(t^{2/3}\), while the fluctuations of \(\log |G|\) scale as \(t^{1/3}\).  We verify these predictions numerically using tensor network simulations of individual circuits. 

We then turn to the weak noise limit, where coherent motion is coupled to slow diffusive modes. 
For simplicity, we derive the large-scale behavior of a partially annealed proxy for the Green's function, while retaining the exact hopping rate disorder. Its large-scale behavior is that of a directed polymer in a random medium~\cite{KARDAR1987582,PhysRevB.43.10728,PhysRevLett.58.2087,PhysRevLett.55.2923,huse1985huse,Bertini1995} (DPRM), with a weak-noise crossover time \(\Ord(\gamma^{-3/2})\). 
(A full treatment of the typical dynamics of Green's functions, including interference effects, is given in the Supplementary Material using replicas~\cite{suppmat}. It yields a DWRM description with the same KPZ scaling and crossover time.)

\textbf{\textit{Particle-conserving random circuits}}---We consider a brickwork circuit on a one-dimensional lattice of qubits, with two-site unitary gates $u_{x,x+1;t}$, acting on alternating bonds at successive time steps. Each gate is drawn independently from the Haar measure subject to particle-number conservation, $[u_{x,x+1;t},n_x+n_{x+1}]=0$, with $n_x=(\mathds{1}+\sigma_x^z)/{2}$.

After every unitary layer, each site is coupled at rate $\gamma$ to a bath at filling \(\nu\). For off-diagonal operators, $\mathcal N_{\nu,\gamma}(\sigma^\pm)=e^{-\gamma}\sigma^\pm$; for diagonal operators $P^{\uparrow/\downarrow}$,
\begin{equation}
  \mathcal N_{\nu,\gamma}(P^a)
  =
  e^{-\gamma}P^a+(1-e^{-\gamma})(\nu P^\uparrow + (1-\nu)P^\downarrow).
  \nonumber
\end{equation}
At $\nu=1/2$, this is simply depolarizing noise. 

The noise is \emph{weakly symmetric}~\cite{Buca_2012,PhysRevA.89.022118,PhysRevB.110.155150,ziereis2025}: although it does not conserve particle number, it preserves \emph{operator charge}. An operator $O$ with charge $q$ satisfies $[\sum_x n_x,O]=q\, O$. Both the unitary gates and the noise channel preserve each operator-charge sector. Note that a particle creation operator $\sigma^+$, which creates a local coherence between number sectors, has $q(\sigma^+)=+1$, while all hydrodynamic observables, e.g., densities, have $q=0$. 

For a fixed circuit realization $U$ and noise $\gamma$, we study the infinite-temperature single-particle Green's function
\begin{equation}
G_\gamma(x,t)
\equiv
\left\langle \sigma_x^-(0)\sigma_0^+(t)\right\rangle_\gamma.
\label{eq:greens-function}
\end{equation}

\textbf{\textit{Strong noise}}---We first consider the
strong-noise limit \(\gamma\gg1\)~\footnote{There is no Zeno effect here. The discrete-time dynamics has \(\Ord(1)\) hop amplitudes even as \(\gamma\to\infty\).}, taking the noise to be depolarizing, \(\nu=1/2\). Since operator charge is conserved, every history contributing to \(G_\gamma\) contains an ever-present \(\sigma^+\), with an associated per-step damping factor $e^{-\gamma}$. Removing this common decay, additional nonidentity operators are suppressed by further powers of \(e^{-\gamma}\). Large $\gamma$ therefore collapses the operator histories onto single-particle dynamics for the position of the coherence, $|G_\gamma(t)\rangle=\sum_x G_\gamma(x,t)|x\rangle$, where $|x\rangle\equiv\sigma_x^+$. Over a half-layer,
\begin{equation}
  |G_\gamma(t+\tfrac12)\rangle=T_t|G_\gamma(t)\rangle,
  \ \
  [T_t]_{xy}=e^{-\gamma}\langle x|U_t\otimes U_t^\ast|y\rangle .
  \label{eq:strong_noise_transfer}
\end{equation}
For an even (odd) half-layer, \(T_t\) is a direct sum of random \(2\times2\) matrices acting on \(\{|x\rangle,|x+1\rangle\}\), with \(x\) even (odd).

Although the unitary evolution preserves the norm in the full operator space, its restriction to the single-coherence subspace does not. A gate can transfer weight from a one-site operator into operators of support two. Under the subsequent depolarizing layer these components are discarded at $\gamma \gg 1$. The matrix $[T_t]_{xx'}$ is consequently contractive, with a gate-dependent loss rate. Intuitively, gates close to the identity generate little operator growth and hence little additional damping, whereas typical gates tend to extend an operator's support.

Equation~\eqref{eq:strong_noise_transfer} therefore describes a discrete-time non-Hermitian Schr\"odinger evolution with random complex hopping amplitudes and complex local attenuation.
On long wavelengths, this gives
\begin{equation}
  \partial_t G_\gamma
  =
  D\partial_x^2 G_\gamma
  +
  [-\Gamma+\eta(x,t)] G_\gamma ,
  \label{eq:directed_wave_main}
\end{equation}
where \(\Gamma=2\gamma+\Ord(1)\) is a uniform decay rate and \(\eta\) is a zero-mean complex short-range random potential with finite variance (see Supplemental Material~\cite{suppmat} for details).
Equation~\eqref{eq:directed_wave_main} describes a DWRM~\cite{PhysRevLett.62.941,PhysRevLett.66.2176,PhysRevA.45.8859,BLUM1992588,1994cond.mat.11022K,PhysRevB.83.052405}: unlike directed polymers, different paths can interfere. 

Define $Z(t) \equiv\sum_x |G_\gamma(x,t)|^2$, $F(t)\equiv-\log Z(t)$, and the distribution $p(x,t)\propto{|G_\gamma (x,t)|^2}$. For each circuit we denote averages with respect to \(p\) by \(\langle \bullet \rangle\), and disorder averages by \(\overline{(\bullet)}\). The DWRM mapping predicts
\begin{equation}
  {\rm Var}[{F(t)}]\sim t^{2/3},
  \quad
  \overline{\langle {x}\rangle^2}\sim t^{4/3}, \quad \overline{\langle x^2\rangle-\langle x\rangle^2}\sim t.
  \label{eq:directed_wave_scaling_main}
\end{equation}
Sample-to-sample wandering of the center of the distribution \(p\) is measured by \(\overline{\langle x\rangle^2}\), and has KPZ scaling, while \(\overline{\langle x^2\rangle-\langle x\rangle^2}\) measures the width of \(p(x,t)\) within a fixed sample, and is diffusive at the DWRM fixed point~\cite{PhysRevLett.62.941}.

While the strong-noise limit is especially simple---the dynamics closes on a single \(\sigma_x^+\)---the same universality is expected at generic noise strength and for non-conserving gates. In each case noise ensures that the relevant operator histories are strings of finite length. These histories interfere, and fluctuations in the gate-dependent
operator-growth rate generate an effective random energy. (A related problem was studied in Ref.~\cite{Nahum_22} for noiseless non-conserving random circuits, where KPZ scaling of local correlators was also anticipated.)

\begin{figure}
    \centering
    \includegraphics[width=1.0\linewidth]{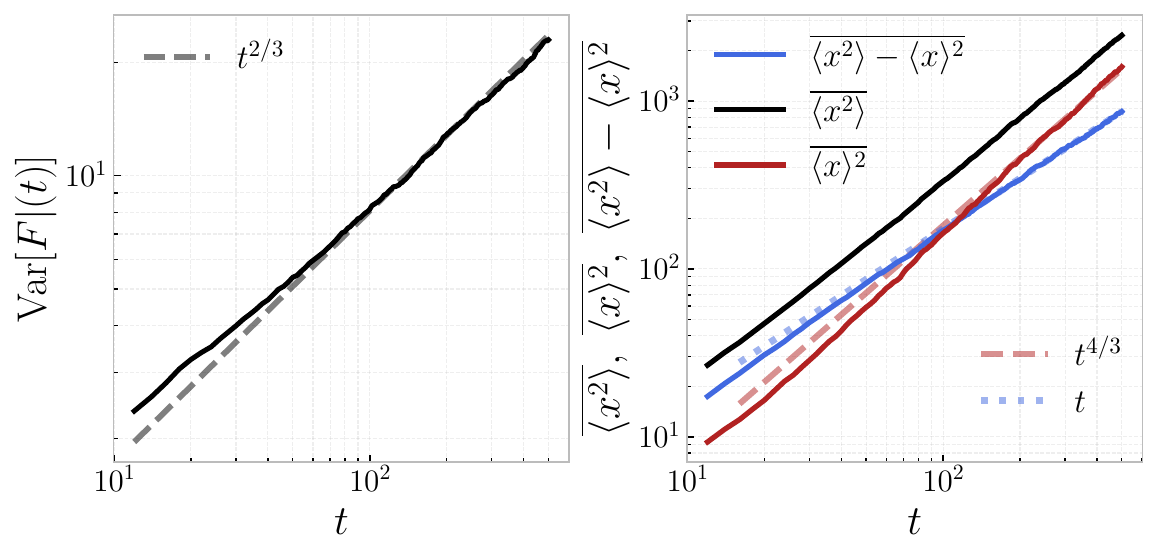}
    \caption{{\textbf{KPZ scaling of Green's functions at strong noise}}. TEBD simulations of noisy operator dynamics at \(\gamma\approx 0.667\), averaged over \(N\approx5000\) circuit samples, showing the expected KPZ scaling exponents in Eq.~\eqref{eq:directed_wave_scaling_main} (dashed guides).}
    \label{fig1:directed_waves}
\end{figure}

We confirm the predicted scalings in matrix-product-state simulations of individual \(U(1)\)-conserving circuits at moderate noise strength \(\gamma=2/3\), as shown in Fig.~\ref{fig1:directed_waves}.

\textbf{\textit{Effective dynamics at weak noise}}---
Ref.~\cite{previous} studied annealed two-copy observables at weak noise using a projected operator-dynamics, or ``source-manifold'', ansatz.
Analytical predictions obtained from this ansatz  were benchmarked against direct numerical simulations of the full two-copy transfer-matrix dynamics and showed excellent agreement in the weak-noise regime. We now extend this framework to study Green's functions at weak noise $\gamma \ll 1$ with spacetime-dependent hopping disorder.

In the neutral sector \(q=0\), an initially diagonal operator, projected back onto the diagonal operator subspace, is described at long wavelengths by the hydrodynamics of a symmetric exclusion process (SEP)~\cite{Rakovszky2018,Khemani_2018,McCulloch2023}. Our principal assumption is that the \(q=1\) sector admits an analogous long-wavelength description after restricting to operators with one local coherence \(\sigma^+\) and diagonal operators elsewhere. The local \(U(1)\)-conserving gates move the coherence microscopically with gate-dependent hopping probabilities and phases, while the surrounding diagonal environment evolves, after projection back to the diagonal subspace, as an SEP fluid with the same local hopping disorder inherited from the gates.



Unlike in Ref.~\cite{previous}, we retain the gate-dependent hopping probabilities and phases explicitly, which are crucial in determining the long-time dynamics of \(G_\gamma\).

The effective dynamics therefore consists of a trajectory \(X\) for the coherence and a spacetime history \(\mathbf n=\{n_x(t)\in\{0,1\}\}_{x\neq X(t)}\) specifying the diagonal operator on every other site.  For a fixed coherence path \(X\), \(\cP_U[\mathbf n|X]\) is the probability of the fluid history \(\mathbf n\) under SEP with the random hopping rates supplied by the gates.  Given this fluid history, the coherence trajectory carries a path amplitude \(\cA_U[X|\mathbf n]\), which contains the same hopping disorder, and also a random phase accumulated along \(X\). With this effective description, the single-particle Green's function can be written as
\begin{equation}
  G_\gamma(x_f,t_f;x_i,t_i)
  \simeq
  \sum_{X:x_i\to x_f}\sum_{\mathbf n}
  \cP_U[\mathbf n|X]\,
  \cA_U[X|\mathbf n] .
  \label{eq:projected_G_path_sum_main}
\end{equation}

Since the phase average of \(G_\gamma\) itself vanishes, the simplest nonzero object is the phase-annealed norm $\mathcal N\equiv E_{\phi}\,|G_\gamma|^2$,
where the average is over phase degrees of freedom of the gates, at fixed microscopic hopping probabilities \(p_j\equiv |\bra{\downarrow\uparrow}u_j\ket{\uparrow\downarrow}|^2\) for a bond $j$. 
By Haar invariance, a \(U(1)\)-symmetric gate may be dressed by independent single-site phase rotations without changing $p_j$. A coherence \(\sigma^+\) thus acquires a random phase, while diagonal operators are invariant. Writing \(X\) and \(\bar X\) for the coherence paths in \(G_\gamma\) and \(G_\gamma^\ast\), phase averaging forces $X=\bar X$, which subsequently undergoes a classical random walk with hopping probabilities \(p_j\)~\cite{von_Keyserlingk_2022,Rakovszky2018,Khemani_2018,Ahlbrecht2012,Joye2011}.

There is additional phase freedom which gives a second constraint. If the coherence is on the left site of a bond, the two possible occupations on the right site correspond to operators $\sigma^+_x (1-n)_{x+1}=\ket{\uparrow\downarrow}\bra{\downarrow\downarrow}$ and $\sigma^+_x n_{x+1}=\ket{\uparrow\uparrow}\bra{\downarrow\uparrow}$.
These operators connect different pairs of \(U(1)\) blocks 
and acquire independent phases.  Therefore, a history survives phase averaging only if the neighboring occupation agrees in the two copies:
\begin{equation}
  n\boxtimes (1-n),\,
  (1-n)\boxtimes n
  \longrightarrow 0 .
  \label{eq:local_agreement_filter_up_down}
\end{equation}

Putting these ingredients together, the phase-annealed norm at fixed real hopping disorder can be written as
\begin{equation}
  \mathcal N(x_f,t_f;x_i,t_i)
  \simeq
  \sum_{X:x_i\to x_f}
  \mathbb P^{\rm coh}[X]\,
  \mathcal Z^{2{\rm -copy}}[X] ,
  \label{eq:locked_norm_two_copy_filter_main}
\end{equation}
where $\mathbb P^{\rm coh}[X]=\prod_t p_{j_t}(X_{t+1}\!\mid X_t)$ is the (classical) probability of the coherence trajectory $X$. 
The factor \(\mathcal Z^{2{\rm -copy}}\) is the probability that two independent exclusion processes, conditioned on the embedded coherence trajectory $X$, have equal occupations on each site neighbouring \(X\).
Without this condition, Eq.~\eqref{eq:locked_norm_two_copy_filter_main} would describe an ordinary stochastic process: a classical walker \(X\), with two independent SEP copies. Eq.~\eqref{eq:local_agreement_filter_up_down} makes the process sub-stochastic by removing incompatible histories.

\textbf{\textit{MFT formulation}}---The surviving histories in the sub-stochastic process form a quasi-stationary ensemble (QSE) on the bath timescale \(\gamma^{-1}\)~\cite{previous}. If the fluid copies have \(\Ord(1)\) entropy density next to the coherence, disagreements occur at an \(\Ord(1)\) rate.  The QSE therefore suppresses either particles or holes at the coherence.  We call the resulting low-entropy region a \emph{void}. We choose a partially polarized bath at density $\nu = \Ord(1)<1/2$, making the particle void the least costly~\footnote{At \(\nu=1/2\), particle and hole voids are degenerate. Once a void has formed, however, its polarization switches only on a nonperturbatively long timescale \(\Ord(\exp(c/\gamma))\)~\cite{previous}, much longer than the timescales algebraic in \(1/\gamma\) considered here. We may therefore regard the void polarization as fixed; choosing \(\nu<1/2\) is simply a convenient choice to break the degeneracy.}.

It is useful to introduce the symbols
\begin{equation}
  \dsym\equiv 1-n,\qquad
  \usym\equiv n,\qquad
  \rsym\equiv (1-\nu)\dsym+\nu\usym,
  \nonumber
\end{equation}
where \(\rsym\) is the local state produced by a bath reset event, at rate \(\gamma\).  In the bulk these three symbols undergo excluded hopping with shared quenched random hopping rates $p_j$. The condition in Eq.~\eqref{eq:local_agreement_filter_up_down} acts on $\rsym$ as
\begin{align}
  \rsym\boxtimes\dsym,\ \dsym\boxtimes\rsym
  &\longrightarrow
  (1-\nu)\,\dsym\boxtimes\dsym,
  \nonumber\\
  \rsym\boxtimes\usym,\ \usym\boxtimes\rsym
  &\longrightarrow
  \nu\,\usym\boxtimes\usym,
  \nonumber\\
  \rsym\boxtimes\rsym
  &\longrightarrow
  (1-\nu)^2\,\dsym\boxtimes\dsym
  +
  \nu^2\,\usym\boxtimes\usym .
  \label{eq:reset_filter_rules_main}
\end{align}

We first keep only the \(\dsym\boxtimes\dsym\) outputs, which restore the particle void.  With this restriction the dynamics closes on the two symbols \(\{\dsym,\rsym\}\): bath events reset the local state to \(\rsym\), while an \(\rsym\) reaching the coherence is replaced by \(\dsym\). Thus the coherence acts, in each copy, as a point absorber for \(\rsym\) symbols,
\begin{equation}
  \rsym\longrightarrow (1-\nu)\dsym,
  \label{eq:no_hazard_rule_main}
\end{equation}
with multiplicative weight \(1-\nu\) for each absorption.

The point-absorber problem gives a quasi-stationary void with parametrically small $\rsym$ density near the coherence, \(\varrho_{0}\ll1\).  The configurations in Eq.~\eqref{eq:reset_filter_rules_main} that produce \(\usym\boxtimes\usym\) are therefore rarely encountered. Moreover, once such a pair is produced, if either returns to the coherence before being reset, it typically does so with a \(\dsym\) symbol in the other copy (since $\varrho_{0}\ll 1$). Thus a history survives only if both \(\usym\)'s avoid returning. In one dimension this no-return conditioning gives an additional weak-noise suppression. The expected number of \(\usym\)'s in the QSE is \(\Ord(\gamma\log(1/\gamma))\) (see End Matter), justifying the point-absorber approximation in Eq.~\eqref{eq:no_hazard_rule_main}.


Under this approximation, the two-copy partition sum factorizes into two identical current-tilted partition sums,
\begin{equation}
  \mathcal Z^{2{\rm -copy}}_{U}
  \simeq
  Z[X]^2,
  \quad
  Z[X]
  =
  \left\langle e^{-sJ_X(T)}\right\rangle_{\{\dsym,\rsym\}},
  \label{eq:current_tilt_fixed_D_main}
\end{equation}
where $e^{-s}=1-\nu$ (i.e., $s = \Ord(1)$) and \(J_X(T)\) is the integrated number of \(\rsym\) symbols absorbed by the coherence. The expectation $\langle \bullet \rangle_{\{\dsym,\rsym\}}$ is with respect to the untilted \(\{\dsym, \rsym\}\)-symbol SEP, with hopping disorder, bulk $\dsym \to \rsym$ resets at rate \(\gamma\), and an absorbing sink at $X$. 

For a given \(X(t)\), this tilted SEP measure admits a Macroscopic Fluctuation Theory (MFT) form~\cite{Bertini_2015,Derrida_2025,PhysRevLett.129.040601,2026arXiv260102319S,10.21468/SciPostPhys.17.2.033},
\begin{equation}
  Z[X]
  =
  \int_{\varrho(X(t),t)=0}
  \mathcal D\varrho\,\mathcal D\pi\,
  \exp\!\left[-S_{\rm MFT}[\varrho,\pi;X,D]\right],
  \label{eq:MFT_representation_main}
\end{equation}
where $D(x,t)$ encodes the microscopic hopping disorder, and where the absorbing boundary condition \(\varrho(X(t),t)=0\) implements the perfect sink in Eq.~\eqref{eq:no_hazard_rule_main}.  The action is
\begin{align}
  S_{\rm MFT}
  =
  \int\! dt\,dx
  &\big[
    \pi\,\partial_t\varrho
    +
    D(x,t)\!
    \left(
      \partial_x\pi\,\partial_x\varrho
      -
      \sigma(\varrho)(\partial_x\pi)^2
    \right)\nonumber\\
    &+
    \gamma\,H_{\rm bath}(\varrho,\pi)
  \big]
  +
  s\,J_X[\varrho,\pi;D],
  \label{eq:MFT_action_main}
\end{align}
with $\sigma(\varrho)=\varrho(1-\varrho)$. Here, \(\varrho\) is the density of \(\rsym\) symbols, \(\pi\) is the corresponding response field and \(H_{\rm bath}\) describes bulk resets. 
For \(D(x,t)\to D_0\), this is the same action previously obtained in Ref.~\cite{previous} from a fully annealed two-replica statistical mechanics model.

To determine the leading-order density profile, we set a uniform diffusivity \(D_0\), and \(X(t)=0\), following Ref.~\cite{previous}. The clean problem has a stationary saddle with \(Z_{D_0}[0]\asymp \exp[-\lambda_0 T]\). The optimal void size $\xi$ is fixed by matching the diffusive relaxation time \(\tau_{\rm void}\sim \xi^2/D_0\) to the bath timescale \(\gamma^{-1}\), while the on-shell action yields $\lambda_0$,
\begin{equation}
  \xi\sim\sqrt{D_0/\gamma},
  \qquad
  \tvoid\sim\gamma^{-1},
  \qquad
  \lambda_0\sim\sqrt{D_0\gamma}.
  \label{eq:void_static_scalings_main}
\end{equation}

The same MFT problem can be used to determine the large-deviation cost of rigidly translating the void.  Setting \(X(t)=vt\), the saddle is stationary in the co-moving frame, and the small-\(v\) expansion gives a correction to the large deviation rate function,
\begin{equation}
  \lambda_v-\lambda_0
  \approx
  v^2\bigg[\frac{1}{4D_0}
  +
  \frac{c}{\sqrt{D_0\gamma}}\bigg] \xRightarrow{\gamma\ll 1} D_{\rm void}\sim\sqrt{D_0\gamma},
\label{eq:polaron-diffusion-constant}
\end{equation}
where $c=\Ord(1)$ and $D_{\rm void}$ is the diffusion constant of the void. 
The void is therefore much slower than the bare coherence. During one void relaxation time it moves only $\Delta X\sim\sqrt{D_{\rm void}\tau_{\rm void}}\sim \gamma^{-1/4}$, i.e., $\Delta X/\xi\sim\gamma^{1/4}\ll1$. (See End Matter for derivations of Eq's~\eqref{eq:void_static_scalings_main}--\eqref{eq:polaron-diffusion-constant}.) 

The stationary saddle analysis above assumes that the coherence position \(X(t)\) can be treated as static on the void-relaxation time.  The self-consistency condition is therefore that, during a time \(\tau_{\rm void}\), the coherence explores a distance much smaller than the void size \(\xi\). This can be verified using a Feynman-polaron ansatz~\cite{previous,FeynmanPolarCrystal}: the internal coherence motion is confined to \(\ell\sim (D_0/\gamma)^{1/3}\ll \xi\). Together with \(\sqrt{D_{\rm void}\tau_{\rm void}}/\xi\sim(\gamma/D_0)^{1/4}\ll1\), this justifies determining the quasi-stationary profile using a static coherence.

After coarse-graining, we are left with an effective description for the slow coordinate \(y(t)\) of the void center,
\begin{equation}
  \mathcal N_{D_0}
  \simeq
  \int \mathcal D y\,
  \exp\left\{
    -\int dt\,
    \left[
      \frac{\dot y^2}{4D_{\rm void}}
      +\Lambda_0
    \right]\right\},
  \label{eq:void_path_integral_main}
\end{equation}
where \(\Lambda_0=2\lambda_0+\delta\lambda_{\rm int}\).  Here \(2\lambda_0\) is themtwo-copy cost of the quasi-stationary void, while \(\delta\lambda_{\rm int}\) is a contribution from integrating out the internal coherence motion. (See End Matter for details.) 

\textbf{\textit{Restoring transport disorder}}---We now restore the hopping disorder \(D(x,t)=D_0+\delta D(x,t)\). 
The fluctuations in the kinetic term in Eq.~\eqref{eq:void_path_integral_main} are irrelevant. The leading effect is the now fluctuating decay rate $\lambda_0 \to \lambda[D;y]$. For a void centered at \(y\), the leading correction to the on-shell MFT action is
\begin{equation}
  V_\xi(y,t)
  \equiv
  2\! \int dx
  \left.
  \frac{\delta \lambda[D;y]}{\delta D(x)}
  \right|_{D_0}
  \!\delta D(x,t)
  \equiv
  \left(K_\xi \star \delta D\right)(y,t),
  \label{eq:random-potential}
\end{equation}
where \(K_\xi\) is fixed by the static-void saddle point and has support on the void scale \(\xi\).  Disorder also perturbs the internal coherence motion, producing an additional random free-energy contribution.  This contribution is parametrically smaller at weak noise (see End Matter).

\begin{figure}[!t]
    \includegraphics[width=1.0\linewidth]{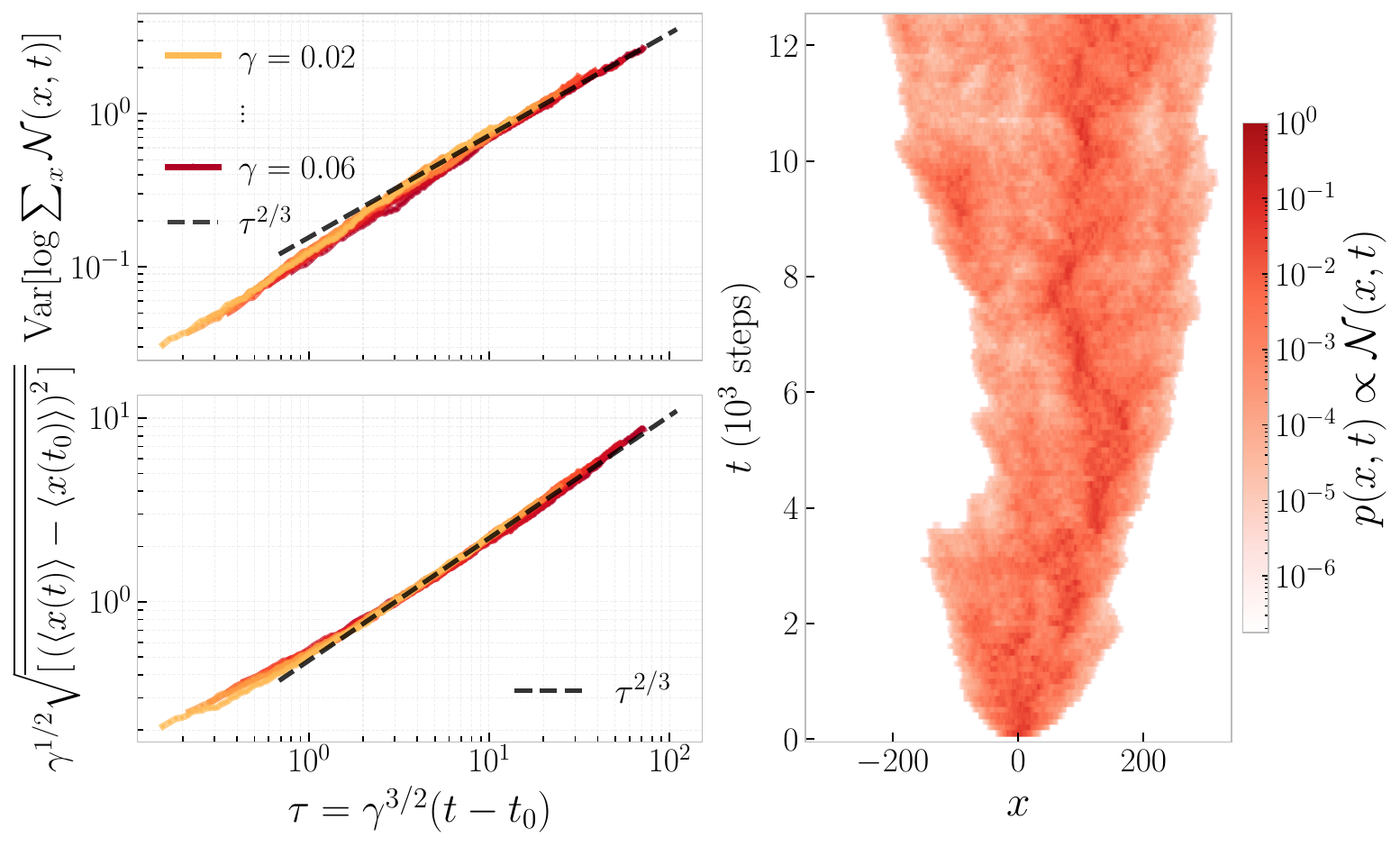}
    \caption{
\textbf{Weak-noise scaling of phase-annealed norm}
\(\mathcal N\). TEBD simulations of the full two-copy transfer matrix dynamics, for bath density \(\nu=1/3\) and coupling strengths \(\gamma\in[0.02,0.06]\), using \(N=1500\) disorder samples. Left: free-energy fluctuations and endpoint wandering with growth consistent with the KPZ exponent \(1/z=2/3\) (guides).  Right: representative endpoint distribution \(p(x,t)\propto \mathcal N(x,t)\) in a single disorder sample.
}
    \label{fig:weak_noise}
\end{figure}

Thus, to leading order, \(\mathcal N\) is described by an effective coarse-grained polymer partition sum. Up to constants,
\begin{equation}
  \mathcal N
  \propto
  \int \mathcal Dy\,
  \exp\left\{
    -\int dt\,
    \left[
      \frac{\dot y^2}{4D_{\rm void}}
      +V_\xi(y,t)
    \right]\right\}.
    \label{eq:weak-noise-DPRE}
\end{equation}

The normalization of \(K_\xi\) follows from the response to a small uniform shift of the diffusivity.  Since \(\lambda_0\sim\sqrt{D_0\gamma}\), a uniform shift \(D_0\to D_0+\delta D\) gives \(\delta\lambda_0\sim \delta D/\xi\).
Thus \(\int dx\,K_\xi(x)\sim \xi^{-1}\), and \(K_\xi(x)\sim \xi^{-2}\).

The natural coarse-grained variables are set by the void width \(\xi\) and the time to diffuse over a distance \(\xi\),
\begin{equation}
  x=\xi x',\quad
  t=\widetilde\tau t',
  \quad
  \widetilde\tau \equiv {\xi^2}/{D_{\rm void}}
  \sim\gamma^{-3/2}.
  \label{eq:weak-noise-scaling1}
\end{equation}
Since \(\int dt\,{\dot y^2}/{D_{\rm void}}=\int d t'\,{(\partial_{t'} y')^2}\), the void diffusion constant is \(\Ord(1)\) in the rescaled coordinates.

We next estimate the random free-energy fluctuation accumulated in one coarse-grained spacetime block. The contribution \(V_\xi\) samples short-range correlated disorder with $\Ord(1)$ variance over a spatial range \(\xi\), with typical response \(K_\xi\sim\xi^{-2}\).  During one rescaled time step, the block therefore contains \(N_\xi\sim \xi\widetilde\tau\sim \xi^4\) independent contributions. The central-limit estimate thus gives
\begin{equation}
  \delta F_\xi
  \sim
  K_\xi \sqrt{N_\xi}
  \sim \Ord(1).
  \label{eq:weak-noise-scaling2}
\end{equation}
The leading coarse-grained theory is therefore a directed polymer with $\Ord(1)$ diffusivity and $\Ord(1)$ random potential in the rescaled variables. The crossover time is parametrically longer than the void-formation time \(\widetilde\tau\gg \tau_{\rm void}\). 
Determining the quasi-stationary void profile at uniform diffusivity \(D_0\) is therefore self-consistent.


We test the weak-noise scaling in Eqs.~\eqref{eq:weak-noise-DPRE}--\eqref{eq:weak-noise-scaling2} by simulating the full two-copy transfer-matrix dynamics for \(\mathcal N\)~\cite{Khemani_2018,Rakovszky2018} using the time-evolving block decimation algorithm~\cite{PhysRevLett.93.040502, PhysRevLett.93.207204,PhysRevLett.93.076401,tenpy2024}. 
For noisy two-copy dynamics, this can be done essentially exactly, to numerical precision~\footnote{Both external noise and phase annealing substantially reduce entanglement growth in the two-copy transfer matrix dynamics.}.

A representative sample is shown in Fig.~\ref{fig:weak_noise} (right). To suppress short-time transients, we measure time relative to a reference \(t_0\gtrsim\tau_{\rm void}\). Figure~\ref{fig:weak_noise} (left) shows that the variance of \(\log\sum_x \mathcal N(x,t)\) and the rescaled wandering length collapse when plotted against \(\tau=\gamma^{3/2}(t-t_0)\), consistent with the predicted weak-noise crossover, while their growth is consistent with KPZ/DPRM universality. The numerical results therefore provide a direct benchmark of both the projected-dynamics ansatz and the weak-noise scaling.

The calculation and numerics above treat the phase-annealed norm \(\mathcal N\), for which the coarse-grained theory is a DPRM. For a typical Green's function, the remaining quenched phase disorder produces interference between coarse-grained paths. A replica treatment of Eq.~\eqref{eq:projected_G_path_sum_main} yields a DWRM description with the same weak-noise scale \(\widetilde\tau\sim\gamma^{-3/2}\) and the same KPZ exponents (see the Supplemental Material~\cite{suppmat}).

\textbf{\textit{Discussion}}---For classical densities and currents in diffusive quantum systems, microscopic hopping disorder is irrelevant at the diffusive fixed point. We have shown that, for intrinsically quantum, phase-sensitive observables, the same disorder can instead be relevant. At weak noise, our analysis uses a physically motivated ansatz for the dynamics of local quantum coherence,  which we tested against direct numerical simulations of the full two-replica transfer-matrix dynamics.
Within this description, the fully annealed Green's-function problem maps onto a clean directed-polymer action. Retaining hopping disorder gives a directed polymer in a random medium for the phase-annealed norm, while restoring the quenched phase disorder gives a directed wave in a random medium. Thus quantum and classical correlators in the same system can be controlled by different universal fixed points.


Several open directions remain. One outstanding task is to verify the weak-noise DWRM scaling directly for genuine single-sample Green's functions, rather than for the phase-annealed norm \(\mathcal N\). This is challenging because of the parametrically large crossover time \(\Ord(\gamma^{-3/2})\) and competing preasymptotic contributions. It would also be valuable to extend the analysis to higher dimensions, where both void formation and strong-disorder fluctuations qualitatively change, and to higher-order phase-sensitive observables, such as multiparticle Green's functions. Recent work~\cite{2026arXiv260728255B} has studied replica moments of equal-time coherence (e.g., $\langle \sigma_x^+(t)\sigma_y^-(t)\rangle$) in driven steady states of a closely related $U(1)$-conserving Brownian-Hamiltonian model. In view of our findings, it would also be interesting to characterize these correlators in individual realizations.

{\textit{Acknowledgments}}---The author thanks Adam Nahum for helpful discussions. I am grateful to J. A. Jacoby, Curt von Keyserlingk, and Sarang Gopalakrishnan for collaboration on earlier work. 
EM is supported by the European Union (ERC, STAQQ, 101171399). Views and opinions expressed are however those of the authors only and do not necessarily reflect those of the European Union or the European Research Council Executive Agency. Neither the European Union nor the granting authority can be held responsible for them.

\bibliography{Refs}

\clearpage
\twocolumngrid
\begin{center}
    \textbf{\large End Matter}
\end{center}

\textbf{\textit{Point-absorber approximation}}---We now check the validity of the point-absorber approximation Eq.~\eqref{eq:no_hazard_rule_main}, by estimating the number of \(\usym\) symbols in the quasi-stationary state. In the static void, the reset-symbol density vanishes at the sink and varies over the scale \(\xi\), so the density near the sink is $\varrho_{\rm core}\sim \xi^{-1}\sim \sqrt{\gamma/D_0}$. The \(\usym\boxtimes\usym\) terms in Eq.~\eqref{eq:reset_filter_rules_main} are therefore generated at rate $\Gamma_{\usym}\sim \varrho_{\rm core}^2\sim \gamma/D_0$, up to nonuniversal factors.

A naive estimate of the number of \(\usym\boxtimes\usym\) pairs in the quasi-stationary state would multiply this production rate by the reset time \(\gamma^{-1}\), since bath resets remove \(\usym\) symbols on that timescale.
This overestimates their weight in the quasi-stationary state. A \(\usym\boxtimes\usym\) pair created near the coherence typically returns to the absorbing point before either symbol is reset. On return, it is paired with a \(\dsym\) symbol in the other copy with high probability (since $\varrho_{\rm core}\ll 1$) and the history is removed by the agreement constraint Eq.~\eqref{eq:local_agreement_filter_up_down}.  

The relevant residence time is therefore a weighted sum over lifetimes, ${\tau_{\usym}^{\rm eff}\sim\int_1^{1/\gamma} dt\, P_{\rm surv}(t)}$, where \(P_{\rm surv}(t)\) is the probability that both \(\usym\) symbols avoid returning to the coherence up to time \(t\). For one diffusive particle initially near an absorbing point, the no-return probability scales as \(t^{-1/2}\); for the pair it scales as \(t^{-1}\). Hence, $\tau_{\usym}^{\rm eff}\sim \int^{1/\gamma} dt\, t^{-1}\sim \log(1/\gamma)$.

The expected number of \(\usym\) symbols present in the quasi-stationary state of the extended \(\{\dsym,\rsym,\usym\}\) substochastic exclusion process is therefore
\begin{equation}
  N_{\usym}^{\rm QSS}
  \sim
  \Gamma_{\usym}\tau_{\usym}^{\rm eff}
  \sim
  \gamma\log(1/\gamma).
  \nonumber
\end{equation}
This is parametrically small at weak noise,
justifying the point-absorber approximation.

\textbf{\textit{MFT stationary saddle equations}}---We seek the stationary saddle equations of the action in Eq.~\eqref{eq:MFT_action_main} for a uniformly translating absorber, \(X(t)=vt\).  The static case $X=0$ is the special case \(v=0\). These saddle equations were previously studied in Ref.~\cite{previous}; for completeness, we briefly review the scaling consequences here. 

In the translating frame,
\begin{equation}
    z=x-vt,\quad \varrho(x,t)=\varrho_v(z),\quad \pi(x,t)=\pi_v(z).
    \nonumber
\end{equation}
For a perfect absorber, the left and right halves of the fluid do not exchange particles and may be treated independently.  We therefore focus on the half-line \(z>0\); the \(z<0\) problem is recovered by sending \(v\to -v\). 

The absorbing constraint gives $\varrho_v(0)=0$, while the tilt \(e^{-sJ_X}\) on absorption events is equivalently imposed as a boundary condition on the response field, $\pi_v(0)=-s$. Far from the absorber, the bath imposes the far-field conditions $\varrho_v(\infty)=1$ and $\pi_v(\infty)=0$. Altogether,
\begin{equation}
  \varrho_v(0)=0,\quad
  \pi_v(0)=-s,\quad
  \varrho_v(\infty)=1,\quad
  \pi_v(\infty)=0 .
  \label{eq:mft_halfline_bcs_end}
\end{equation}

For Poisson reset dynamics, the bath Hamiltonian is
\begin{equation}
  H_{\rm bath}(\varrho,\pi)=(1-\varrho)(e^{-\pi}-1),
  \nonumber
\end{equation}
so that at \(\pi=0\) the deterministic equation contains the relaxation term \(\gamma(1-\varrho)\).  Varying Eq.~\eqref{eq:MFT_action_main} then gives
\begin{align}
  -v\varrho_v'
  &=
  D_0\varrho_v''
  -2D_0\partial_z\!\left[\sigma(\varrho_v)\pi_v'\right]
  +\gamma(1-\varrho_v)e^{-\pi_v},
  \nonumber
  \\
  -v\pi_v'
  &=
  -D_0\pi_v''
  -D_0\sigma'(\varrho_v)(\pi_v')^2
  +\gamma(1-e^{-\pi_v}).
  \nonumber
\end{align}
Here primes denote \(z\)-derivatives. Introducing
\begin{equation}
  \xi=\sqrt{D_0/\gamma},\quad
  \zeta=z/\xi,\quad
  u={v\xi}/{D_0}={v}{(D_0\gamma)^{-1/2}},
  \nonumber
\end{equation}
the half-line equations become
\begin{align}
  -u\partial_\zeta\rho
  &=
  \partial_\zeta^2\rho
  -2\partial_\zeta\!\left[\sigma(\rho)\partial_\zeta\pi\right]
  +(1-\rho)e^{-\pi},
  \label{eq:mft_rho_dimensionless_end}
  \\
  -u\partial_\zeta\pi
  &=
  -\partial_\zeta^2\pi
  -\sigma'(\rho)(\partial_\zeta\pi)^2
  +(1-e^{-\pi}),
  \label{eq:mft_pi_dimensionless_end}
\end{align}
with boundary conditions in Eq.~\eqref{eq:mft_halfline_bcs_end}. The left half-line is then obtained by \(u\to -u\) in Eqs~\eqref{eq:mft_rho_dimensionless_end}--\eqref{eq:mft_pi_dimensionless_end}.

This dimensionless form implies a stationary profile width $\xi\sim\sqrt{D_0/\gamma}$, and that quasi-stationary decay rate has the scaling form
\begin{equation}
  \lambda_{\rm fl}(v)
  =
  \sqrt{D_0\gamma}\,
  \hat\lambda_{\rm fl}
  \big({v}/{\sqrt{D_0\gamma}}\big).
  \nonumber
\end{equation}
For \(v=0\), this gives $\lambda_0\sim\sqrt{D_0\gamma}$.

Since the quasi-stationary decay rate combines both left and right half-lines, reflection symmetry is restored, and $\hat\lambda_{\rm fl}$ must be even. The small-\(v\) expansion is thus as given in Eq.~\eqref{eq:polaron-diffusion-constant}.

\textbf{\textit{Void-shape response to hopping disorder}}---We now justify evaluating the leading disorder correction on the clean void saddle. Let \(\Phi=(\varrho,\pi)\) denote the MFT fields, and let \(\Phi_0\) be the stationary saddle at uniform diffusivity \(D_0\).  For a void centered at \(y\), the on-shell decay rate is
\[
  \lambda[D;y]
  =
  \frac{1}{T}
  S_{\rm MFT}[\Phi_*(D);D,y],
\]
where \(\Phi_*(D)\) is the optimal profile for the given diffusivity field.
Writing \(D=D_0+\delta D\), the first-order expansion gives
\[
\begin{split}
  \delta\lambda
  &=
  \frac{1}{T}
  \int dt\,dx\,
  \left.
  \frac{\delta S_{\rm MFT}}{\delta D(x,t)}
  \right|_{\Phi_0,D_0}
  \delta D(x,t)
  \\
  &\quad
  +
  \frac{1}{T}
  \int dt\,dx\,
  \left.
  \frac{\delta S_{\rm MFT}}{\delta\Phi(x,t)}
  \right|_{\Phi_0,D_0}
  \delta\Phi(x,t)
  +O(\delta D^2).
\end{split}
\]
The second line vanishes because \(\Phi_0\) is a saddle.  Thus the linear disorder correction is obtained by differentiating the action explicitly with respect to \(D\), evaluated on the clean void profile.  This yields Eq.~\eqref{eq:random-potential} in the main text.

\textbf{\textit{Feynman--polaron ansatz}}---We next recall how the fast internal motion of the coherence was treated in Ref.~\cite{previous}.  In that work, the point-absorber MFT problem was further simplified by neglecting particle exclusions.  In this independent-particle approximation the tilted absorber problem can be gaussianized by a Cole--Hopf transformation, so that the fluid fields may be integrated out explicitly. This gives a nonlocal effective action for the coherence trajectory \(X(t)\) alone. 
The Feynman-polaron ansatz~\cite{FeynmanPolarCrystal} is a controlled quadratic approximation to this trajectory action: the slowly relaxing deformation of the fluid is represented by a collective coordinate \(y(t)\), while the bare coherence moves relative to it,
\[
  X(t)=y(t)+z(t).
\]
The resulting fully annealed norm is approximated as
\[
  \mathcal N_{D_0}
  \simeq
  \int \mathcal D y\,\mathcal D z\,
  \exp\!\left[-S_{\rm eff}^{(0)}[y,z]\right],
\]
with
\begin{equation}
  S_{\rm eff}^{(0)}[y,z]
  =
  \int dt\left[
    2\lambda_0
    +
    \frac{\dot y^2}{4D_{\rm void}}
    +
    \frac{\dot z^2}{4D_0}
    +
    \frac{D_0}{4\ell^4}z^2
  \right].
  \label{eq:feynman_eff_action_end}
\end{equation}
Here \(2\lambda_0\) is the two-copy cost of the quasi-static void, \(D_{\rm void}\) is the diffusion constant of the slow void coordinate, and \(D_0\) is the bare coherence diffusion constant.  The final term confines the coherence to a variational length \(\ell\) around the void center.

The scaling of \(\ell\) follows from a simple balance.  Confining the bare coherence to width \(\ell\) costs \(D_0/\ell^2\) per unit time.  The rapidly moving coherence acts as a distributed sink over an interval of length \(\ell\) around \(y(t)\). Since this interval is parametrically smaller than the void size, the outer quasi-stationary profile is unchanged to leading order (although shifted $\pm\ell/2$ on each side); the additional fluid cost is local, namely the cost of keeping the swept interval depleted against bath resets.  This costs \(\gamma\ell\) per unit time.  

Minimizing ${D_0}/{\ell^2}+\gamma\ell$ therefore gives
\begin{equation}
  \ell\sim \left({D_0}/{\gamma}\right)^{1/3},
  \qquad
  \tau_\ell\sim {\ell^2}/{D_0}
  \sim D_0^{-1/3}\gamma^{-2/3}.
  \label{eq:inner_length_end}
\end{equation}

Since \(\ell/\xi\sim(\gamma/D_0)^{1/6}\ll1\), the coherence explores only a parametrically small part of the void.  Moreover \(\tau_\ell\ll\tau_{\rm void}\ll\widetilde\tau\), so the internal coordinate equilibrates before the void profile evolves appreciably, and long before hopping disorder produces $\Ord(1)$ coarse-grained free-energy fluctuations. This justifies using the static quasi-stationary void profile when deriving the slow directed-polymer theory for \(y(t)\).

\textbf{\textit{Internal motion with hopping disorder}}---We now estimate the random potential for the slow void coordinate \(y(t)\) generated by restoring hopping disorder in the Feynman--polaron description and integrating out the fast internal coordinate \(z(t)\). We show that this contribution is parametrically smaller than the random potential \(V_\xi(y,t)\), which comes from fluctuations of the static void decay rate \(\lambda_0\) and is retained in the main text.

For fixed \(y(t)\), define the effective action after
integrating out the fast relative coordinate \(z(t)\) by
\vspace{-2mm}
\begin{equation}
  \exp\left[-S_{\rm eff}[y;D]\right]
  =
  \int \mathcal D z\,
  \exp\left[-S_{\rm in}[y,z;D]\right],\nonumber
\end{equation}
where $S_{\rm in}$ includes only the terms in Eq.~\eqref{eq:feynman_eff_action_end} that depend on the internal coordinate $z$. 

Restoring \(D(x,t)=D_0+\delta D(x,t)\), expand
\[
  S_{\rm in}[y,z;D]
  =
  S_{\rm in}^{(0)}[y,z]
  +
  \delta S_{\rm in}[y,z;\delta D]
  +O(\delta D^2),
\]
with
\[
  \delta S_{\rm in}[y,z;\delta D]
  =
  \int dt\,dx\,
  \left.
  \frac{\delta S_{\rm in}}{\delta D(x,t)}
  \right|_{D_0}
  \delta D(x,t).
\]
Therefore,
\begin{align}
  S_{\rm eff}[y;D]-S_{\rm eff}^{(0)}[y]
  &=
  -\log
  \left\langle
  e^{-\delta S_{\rm in}[y,z;\delta D]}
  \right\rangle_{0,z;y}\nonumber\\
  &=
  \left\langle
  \delta S_{\rm in}[y,z;\delta D]
  \right\rangle_{0,z;y}
  +O(\delta D^2),
  \nonumber
\end{align}
where \(\langle\cdots\rangle_{0,z;y}\) denotes the clean bound-state average of the fast coordinate at fixed \(y(t)\).

Thus the first-order correction has the form
\begin{align}
  V_\ell(y,t) &\equiv S_{\rm eff}[y;D]-S_{\rm eff}^{(0)}[y]\nonumber\\
  &= 
  \int dt\,dx\,
  K_\ell(x-y(t))\,\delta D(x,t)
  +\cdots,
\label{eq:inner-motion-correction}
\end{align}
with $K_\ell(x-y)
  \equiv
  \left\langle
  \left.
  (\delta S_{\rm in}/{\delta D(x,t)})
  \right|_{D_0}
  \right\rangle_{0,z;y}$.

The convolution form follows from translation invariance of the clean problem: at \(D_0\), the bound state depends on \(x\) only through \(x-y(t)\).

The kernel is supported on the inner length \(\ell\).  Its total weight is fixed by applying a uniform shift \(D_0\to D_0+\delta D\).  Since the internal confinement energy scale is \(D_0/\ell^2\), one requires $\int dx\,K_\ell(x)\sim \ell^{-2}$, i.e., $K_\ell\sim \ell^{-3}$.

The clean action \(S_{\rm eff}^{(0)}[y]\) already includes the result of integrating out the fast coordinate \(z\) in Eq.~\eqref{eq:feynman_eff_action_end}.  This produces an additional, nonrandom contribution to the decay rate of order the oscillator ground-state energy,
\[
  \lambda_{\rm in}
  \sim
  {D_0}/{\ell^2}
  \sim
  D_0^{1/3}\gamma^{2/3},
\]
where in the last step we used
\(\ell\sim(D_0/\gamma)^{1/3}\). This correction is subleading compared with the quasi-static void cost, $\lambda_{\rm in}/{\lambda_0} \sim ({\gamma}/{D_0})^{1/6}\ll 1$.

We now estimate the fluctuating part of $V_\ell(y,t)$ in Eq.~\eqref{eq:inner-motion-correction}. Over one coarse-grained block of duration $\widetilde\tau=\xi^2/D_{\rm void}\sim\gamma^{-3/2}$, the kernel $K_\ell$ samples \(N_\ell\sim\ell\widetilde\tau\) independent random hoppings (the $\delta D(x,t)$).  Since \(K_\ell\sim\ell^{-3}\), the central-limit estimate gives
\begin{equation}
  \delta F_\ell
  \sim
  K_\ell\sqrt{N_\ell}
  \sim
  \ell^{-3}\sqrt{\ell\tau_\xi}
  \sim
  \gamma^{1/12},
  \label{eq:inner_disorder_scaling_end}
\end{equation}
up to powers of \(D_0\) and nonuniversal constants.

Thus the random potential generated by integrating out the internal coherence motion vanishes in the weak-noise limit on the same coarse-grained blocks where the void-scale random potential \(V_\xi(y,t)\) gives \(\Ord(1)\) fluctuations.
The leading hopping-disorder contribution to the directed-polymer theory is therefore the fluctuation of the quasi-static void decay rate \(\lambda_0\), not the internal coherence motion.

\clearpage

\newpage

\widetext

\makeatletter
\begin{center}
\textbf{\large Supplementary Materials: KPZ Superdiffusion of Local Correlators in Diffusive Random Quantum Circuits}

\vspace{3mm}

Ewan McCulloch

\vspace{2mm}

\textit{\small Laboratoire de Physique de l'École Normale Supérieure, CNRS,\\
ENS \& Université PSL; 24 rue Lhomond, 75005 Paris, France}

\makeatother


\makeatother

\end{center}
\setcounter{equation}{0}
\setcounter{figure}{0}
\setcounter{table}{0}
\setcounter{page}{1}
\makeatletter
\renewcommand{\theequation}{S\arabic{equation}}
\renewcommand{\thefigure}{S\arabic{figure}}

\section{Discrete directed-wave equation at strong noise}
\label{app:strong_noise_dwrm}

We show explicitly that the transfer matrix evolution in Eq.~\eqref{eq:strong_noise_transfer} obtained in the strong-noise limit is a discrete directed wave in a random medium.

\subsubsection{Mapping to discrete directed waves}

Consider a gate acting on a bond \(j=(x,x+1)\). In the sectors with zero and two particles it acts by phases \(e^{i\phi_{0,j}}\) and \(e^{i\phi_{2,j}}\), while its one-particle block may be parameterized as~\cite{TilmaSudarshan2002}
\begin{equation}
  u_j^{(1)}
  =
  e^{i\chi_j}
  \begin{pmatrix}
    e^{i\alpha_j}\sqrt{1-p_j}
    &
    e^{i\beta_j}\sqrt{p_j}
    \\
    -e^{-i\beta_j}\sqrt{p_j}
    &
    e^{-i\alpha_j}\sqrt{1-p_j}
  \end{pmatrix}.
  \label{eq:app_u1_parameterization}
\end{equation}
For the Haar ensemble, $p_j\sim {\rm Unif}[0,1]$, and $\phi_{0,j},\phi_{2,j},\chi_j,\alpha_j,\beta_j\sim {\rm Unif}[0,2\pi)$, with all variables independent.

We project onto the single-coherence basis $\{|L\rangle=\sigma_x^+, |R\rangle=\sigma_{x+1}^+\}$. After removing the common depolarizing factor \(e^{-\gamma}\), the exact projected block ($[T_t]_{xx'}$ in Eq.~\eqref{eq:strong_noise_transfer}) is
\begin{equation}
  M_j
  =
  e^{i\Phi_j}
  \begin{pmatrix}
    \sqrt{1-p_j}\cos\delta_j\,e^{-i\alpha_j}
    &
    -i\sqrt{p_j}\sin\delta_j\,e^{-i\beta_j}
    \\
    i\sqrt{p_j}\sin\delta_j\,e^{i\beta_j}
    &
    \sqrt{1-p_j}\cos\delta_j\,e^{i\alpha_j}
  \end{pmatrix}, \qquad \Phi_j=\frac{\phi_{2,j}-\phi_{0,j}}{2}, \qquad \delta_j
  =
  \chi_j-\frac{\phi_{0,j}+\phi_{2,j}}{2}.
  \label{eq:app_projected_block}
\end{equation}

A local rephasing at the input and output of each half-layer brings this block to
\begin{equation}
  \widetilde M_j
  =
  \begin{pmatrix}
    r_j e^{i\zeta_j} & h_j\\
    h_j & r_j e^{-i\zeta_j}
  \end{pmatrix},
  \label{eq:app_gauge_fixed_block}
\end{equation}
with $h_j=\sqrt{p_j}\,|\sin\delta_j|$ and $r_j=\sqrt{1-p_j}\,|\cos\delta_j|$. The residual phase \(\zeta_j\) is uniform on \([0,2\pi)\) and independent of \(h_j\) and \(r_j\). Explicitly, after fixing the two hopping amplitudes to be positive, the residual phase has the form
\[
  \zeta_j^m
  =
  \frac{\pi}{2}+\beta_j^m-\alpha_j^m
  +\theta_{j,L}^m-\theta_{j,R}^m
  \pmod{2\pi},
\]
up to the signs absorbed into \(h_j\) and \(r_j\). Since \(\beta_j^m-\alpha_j^m\) is a fresh uniform phase for every gate, \(\zeta_j^m\) is uniform and independent of the incoming gauge phases and of \(h_j^m,r_j^m\).

Any signs of \(\sin\delta_j\) and \(\cos\delta_j\) have been absorbed into the rephasing. Thus, on every active spacetime bond,
\begin{equation}
  \begin{pmatrix}
    \psi_{j,L}^{m+1}\\
    \psi_{j,R}^{m+1}
  \end{pmatrix}
  =
  \begin{pmatrix}
    r_j^m e^{i\zeta_j^m} & h_j^m\\
    h_j^m & r_j^m e^{-i\zeta_j^m}
  \end{pmatrix}
  \begin{pmatrix}
    \psi_{j,L}^{m}\\
    \psi_{j,R}^{m}
  \end{pmatrix}.
  \label{eq:app_exact_bond_update}
\end{equation}
This is already a discrete-time directed wave in a random medium: the hopping is real and symmetric, while the diagonal amplitudes are complex and random. Equation~\eqref{eq:app_exact_bond_update} can be rewritten as
\begin{align}
  \psi_{j,L}^{m+1}-\psi_{j,L}^{m}
  &=
  h_j^m
  \left(
    \psi_{j,R}^{m}-\psi_{j,L}^{m}
  \right)
  +
  V_{j,L}^m\psi_{j,L}^{m},
  \nonumber\\
  \psi_{j,R}^{m+1}-\psi_{j,R}^{m}
  &=
  h_j^m
  \left(
    \psi_{j,L}^{m}-\psi_{j,R}^{m}
  \right)
  +
  V_{j,R}^m\psi_{j,R}^{m},
  \label{eq:app_discrete_dwrm}
\end{align}
where $V_{j,L}=h_j-1+r_j e^{i\zeta_j}$, and $V_{j,R}=h_j-1+r_j e^{-i\zeta_j}=V_{j,L}^{\ast}$.

\subsubsection{Noise statistics: means}

For the Haar distribution, we have the following averages,
\begin{equation}
  \overline{h}
  =
  \overline{\sqrt p}\,
  \overline{|\sin\delta|}
  =
  \frac{2}{3}\frac{2}{\pi}
  =
  \frac{4}{3\pi},\qquad
  \overline{h^2}
  =
  \overline p\,
  \overline{\sin^2\delta}
  =
  \frac14,
  \qquad
  \overline{r^2}
  =
  \overline{1-p}\,
  \overline{\cos^2\delta}
  =
  \frac14.
\end{equation}
Because \(\zeta\) is uniform, we have $\overline{r e^{\pm i\zeta}}=0$. Hence $\overline{V_{j,L}}=\overline{V_{j,R}}=\frac{4}{3\pi}-1$.

Define $\delta h_j=h_j-\overline h$ and the centered complex noise
\begin{equation}
  \eta_{j,L}
  \equiv
  V_{j,L}+\lambda_0
  =
  \delta h_j+r_j e^{i\zeta_j},
  \qquad
  \eta_{j,R}
  =
  \delta h_j+r_j e^{-i\zeta_j}
  =
  \eta_{j,L}^{\ast},
  \label{eq:app_centered_noise}
\end{equation}
Furthermore, define
\begin{equation}
    B\equiv\overline{\delta h_j^2}=\frac14-\frac{16}{9\pi^2} \simeq 0.06987, \qquad A\equiv B+\overline{r^2}=\frac12-\frac{16}{9\pi^2}\simeq 0.31987.
\end{equation}

\subsubsection{Noise statistics: covariances}

Let \(s,s'\in\{L,R\}\). The two-point correlation functions of the complex noise $\eta$ (Eq.~\eqref{eq:app_centered_noise}) are
\begin{align}
  \overline{
    \eta_{j,s}^{m}\eta_{j',s'}^{m'}
  }
  =
  \delta_{mm'}\delta_{jj'}\,
  \bigl[\mathcal C_{20}\bigr]_{ss'},
  \qquad
  \overline{
    \eta_{j,s}^{m}\eta_{j',s'}^{m'\ast}
  }
  =
  \delta_{mm'}\delta_{jj'}\,
  \bigl[\mathcal C_{11}\bigr]_{ss'},
  \label{eq:app_exact_noise_correlators}
\end{align}
with covariance matrices
\begin{equation}
  \mathcal C_{20}
  =
  \begin{pmatrix}
    B & A\\
    A & B
  \end{pmatrix},
  \qquad
  \mathcal C_{11}
  =
  \begin{pmatrix}
    A & B\\
    B & A
  \end{pmatrix}.
  \label{eq:app_exact_covariance_matrices}
\end{equation}

The random hopping and multiplicative noise are also correlated:
\begin{equation}
  \overline{
    \delta h_j^m\eta_{j',s}^{m'}
  }
  =
  B\,\delta_{jj'}\delta_{mm'},
  \qquad
  s=L,R.
\end{equation}
These formulas are exact for the discrete time transfer matrix evolution defined in Eq.~\eqref{eq:strong_noise_transfer}. Correlations vanish exactly for distinct half-layers or distinct bonds in the same half-layer. The only spatial correlation is between the two endpoints of the same gate.

\subsubsection{Continuum limit}

On scales large compared with one bond and one half-layer, Eq.~\eqref{eq:app_discrete_dwrm} coarse-grains to
\begin{equation}
  \partial_t\psi
  =
  D\partial_x^2\psi
  +
  [-\Gamma_0+\eta(x,t)]\psi
  +
  \partial_x[
    \delta D(x,t)\partial_x\psi
  ]
  +\ldots .
  \label{eq:app_continuum_random_diffusivity}
\end{equation}
Here \(D\), \(\Gamma_0\), and the continuum noise covariances are nonuniversal but finite and \(\Ord(1)\). The random-diffusivity term contains two spatial derivatives and is irrelevant relative to the multiplicative disorder. Dropping it gives
\begin{equation}
  \partial_t\psi
  =
  D\partial_x^2\psi
  +
  [-\Gamma_0+\eta(x,t)]\psi.
  \label{eq:app_continuum_dwrm}
\end{equation}
Restoring the common factor \(e^{-\gamma}\) per half-step shifts the uniform decay rate \(\Gamma_0\to\Gamma=\Gamma_0+2\gamma\). This gives Eq.~\eqref{eq:directed_wave_main} of the main text, up to the rephasing of \(G_\gamma\), which is left implicit there.

\section{Replica mapping to directed-waves in a random medium at weak noise}
\label{app:weak_noise_dwrm}

Assuming that the projected operator dynamics ansatz captures the long-wavelength single-particle Green’s function at weak noise, this appendix derives its coarse-grained description as a directed wave in a random medium. The projected dynamics is the same source-manifold ansatz introduced and tested for annealed two-replica observables in Ref.~\cite{previous}. 

The leading effect of quenched hopping disorder is discussed in the main text. Here we isolate the additional effects of the random gate phases. We restore $D_U(x,t)=D_0+\delta D_U(x,t)$ at the end and combine the two sources of disorder in the resulting directed-wave description.

\subsection{Effective projected dynamics path sum}
\label{app:replica_operators}

We begin from the effective projected-dynamics ansatz for the single-particle Green function, Eq.~\eqref{eq:projected_G_path_sum_main} in the main text (see also Ref.~\cite{previous}),
\begin{equation}
  G_U(x_f,t_f;x_0,t_0)
  \simeq
  \sum_{X:x_0\to x_f}
  \sum_{\boldsymbol{n}}
  \mathcal P_U[\boldsymbol n|X]\,\mathcal A_U[X|\boldsymbol n].
  \label{eq:app_projected_path_sum}
\end{equation}
In this appendix we keep the circuit realization $U$ explicit as a subscript in the Green's function, with the noise strength $\gamma$ implicit. Here $X(t)$ is the coherence trajectory and $\boldsymbol n$ denotes a microscopic fluid history of the surrounding diagonal operators. We now define the two factors on the right-hand side. 

For a prescribed trajectory $X(t)$, consider the two-site gate on a bond $j=(x,x+1;t)$ visited by the coherence. Within the projected operator subspace, the coherence is always accompanied by a diagonal operator on the neighboring site at this bond. Without loss of generality, suppose that the coherence is on the left site. The operator on this bond is then of the form $\sigma_x^+ P_{x+1}^n$, where $n\in\{0,1\}$ is the occupation on site $x+1$ ($P^0\equiv \ket{0}\bra{0}$ corresponds to an empty site, while $P^1\equiv \ket{1}\bra{1}$ corresponds to an occupied site). A $U(1)$-conserving gate acts within this
subspace as
\begin{equation}
 U_j\,\sigma_x^+ P_{x+1}^n\,U_j^\dagger
 =
 c_{j,\rm s}(n)\,
 \sigma_x^+P_{x+1}^n
 +
 c_{j,\rm h}(n)\,
 P_x^n\sigma_{x+1}^+, \qquad |c_{j,\rm s}(n)|^2+|c_{j,\rm h}(n)|^2=1.
 \label{eq:projected-coherence-update}
\end{equation}
The first term leaves both operators at their initial positions, while the second term swaps them. The second equality follows from unitarity. For a two-qubit $U(1)$-conserving gate the hopping probabilities are independent of $n$, $|c_{j,\rm h}(n)|^2=p_j$, $|c_{j,\rm s}(n)|^2=1-p_j$, where $p_j$ is the random hopping probability associated with the spacetime gate $j$.

For a fixed path $X$ and fluid history $\boldsymbol n$, we therefore define the amplitude
\begin{equation}
 \mathcal A_U[X|\boldsymbol n]
 =
 \prod_{j\in X}
 c_{j,a_j}(n_j) ,
 \label{eq:coherence-path-amplitude}
\end{equation}
where $a_j\in\{{\rm s},{\rm h}\}$ indicate a `stay' or a `hop' at gate $j$ (as prescribed by $X$), and $n_j$ is the neighbouring fluid occupation at the gate $j$. Thus every factor in this amplitude depends on the fluid history sampled along the path.

The amplitude $\mathcal A_U[X|\boldsymbol n]$ contains the factors accumulated at all gates visited by the coherence, for a given occupation history on the neighboring sites.  On the remaining bonds we assume that the dynamics, projected onto diagonal operators, is governed by a symmetric exclusion process with the quenched hopping rates $p_j$. This effective dynamics is therefore represented by a history $\boldsymbol n$ with conditional probability $\mathcal P_U[\boldsymbol n|X]$, where conditioning on $X$ fixes the sequence of `stay' and `hop' moves, and hence the corresponding occupation transfers, on the gates crossed by the coherence.  In the presence of an external bath, $\mathcal P_U[\boldsymbol n|X]$ is the corresponding conditional history probability for the bath-coupled exclusion process. Combining these two factors, $\mathcal A_U[X|\boldsymbol{n}]$ and $\mathcal P_U[\boldsymbol{n}|X]$, and summing over possible trajectories $X$ and occupation histories $\boldsymbol{n}$, gives the effective path sum Eq.~\eqref{eq:app_projected_path_sum} for the single-particle Green's function.

\subsection{Phase-averaged replica statistical mechanics model}

To access typical circuit realizations, we consider the balanced moments
\begin{equation}
  M_{q,U}
  =
  \mathbb{E}_{\phi}\!\left[G_U^q G_U^{*q}\right],
  \label{eq:app_balanced_moment}
\end{equation}
where $\mathbb{E}_{\phi}$ denotes the average over the gate phases at fixed hopping probabilities $\{p_j\}$. (The subscript $U$ in $M_{q,U}$ denotes only the fixed hopping probabilities $\{p_j\}$.) We refer to the $q$ copies of $G_U$ as the \emph{unconjugated replicas}, labelled by $a=1,\ldots,q$, and to the $q$ copies of $G_U^*$ as the \emph{conjugate replicas}, labelled by $b=1,\ldots,q$. Moments containing unequal numbers of $G_U$ and $G_U^*$ vanish under the phase average.

Using Eq.~\eqref{eq:app_projected_path_sum}, the replicated path sum is
\begin{align}
  M_{q,U} \simeq
  \sum_{\{X_a,\boldsymbol n_a\}}
  \sum_{\{\overline X_{a},
                \overline{\boldsymbol n}_{a}\}}
  \mathbb{E}_{\phi}\!\Bigg[
  \prod_{a=1}^{q}
  \mathcal P_U[\boldsymbol n_a|X_a]\,
  \mathcal A_U[X_a|\boldsymbol n_a]
  \prod_{b=1}^{q}
  \mathcal P_U[
      \overline{\boldsymbol n}_{b}
      |\overline X_{b}]
  \mathcal A_U^*[
      \overline X_{b}
      |\overline{\boldsymbol n}_{b}]
  \Bigg].
  \label{eq:app_replicated_path_sum}
\end{align}
Before phase averaging, the path sum therefore contains $q$ unconjugated coherence paths $X_a$, $q$ conjugate coherence paths $\overline X_{b}$, and one fluid history attached to each of these paths.

\subsubsection{Local phase constraints and replica pairing}

As in the $q=1$ calculation in the main text, the phase average imposes two local constraints. The first follows from a sitewise rephasing freedom of the gate ensemble. By Haar invariance, a $U(1)$-symmetric two-qubit gate may be dressed by independent single-site phase rotations without changing either the measure or its hopping probability $p_j$. Under a rotation $V_x(\varphi)=e^{i\varphi n_x}$, a diagonal operator is invariant, whereas $V_x(\varphi)\sigma_x^+V_x^\dagger(\varphi) =e^{i\varphi}\sigma_x^+$. A coherence at site $x$ therefore acquires the phase $e^{i\varphi}$ in an unconjugated replica and the opposite phase $e^{-i\varphi}$ in a conjugate replica.

Denote the number of coherences at the spacetime point $(x,t)$ among the unconjugated and conjugated replicas respectively by
\begin{equation}
  N_{x,t}
  =
  \sum_{a=1}^q
  \mathbf 1\!\left[X_a(t)=x\right],
  \qquad
  \overline N_{x,t}
  =
  \sum_{a=1}^q
  \mathbf 1\!\left[\overline X_{a}(t)=x\right].
\end{equation}
Averaging the local phase gives
\begin{equation}
  \int\frac{d\varphi_{x,t}}{2\pi}\,
  e^{i\varphi_{x,t}(N_{x,t}-\overline N_{x,t})}
  =
  \delta_{N_{x,t},\overline N_{x,t}}.
  \label{eq:app_local_coherence_number_constraint}
\end{equation}
Thus, at every site and time between two-qubit gates, the number of coherences summed over the unconjugated replicas must equal the corresponding number summed over the conjugate replicas.

The second phase constraint involves the occupation of the site neighbouring the coherence at a gate. Fix a spacetime gate $j=(x,x+1;t)$. Let $N^n_{x,j}$ count the unconjugated replicas for which, immediately before the gate, the operator on the bond is $\sigma_x^+P_{x+1}^n$, and let $N^n_{x+1,j}$ count those for which it is $P_x^n\sigma_{x+1}^+$. Here $n\in\{0,1\}$ is the occupation of the site neighbouring the coherence. We define $\overline N^n_{x,j}$ and $\overline N^n_{x+1,j}$ analogously for the conjugate replicas.

The sitewise constraint Eq.~\eqref{eq:app_local_coherence_number_constraint} already implies $N^0_{x,j}+N^1_{x,j} =\overline N^0_{x,j}+\overline N^1_{x,j}$, and similarly for $x\to x+1$. There is also an independent random phase associated with the pair of $U(1)$ charge sectors connected by the coherence. For example, $\sigma_x^+P_{x+1}^0=|10\rangle\langle00|$ connects the charge-zero and charge-one blocks, whereas $\sigma_x^+P_{x+1}^1=|11\rangle\langle01|$ connects the charge-one and charge-two blocks. The corresponding phase depends on the neighbouring occupation $n$, and its average imposes the second constraint
\begin{equation}
  N^n_{x,j}+N^n_{x+1,j}
  =
  \overline N^n_{x,j}+\overline N^n_{x+1,j},
  \qquad n=0,1.
  \label{eq:app_neighbour_counting_constraint}
\end{equation}
For $q=1$, Eqs.~\eqref{eq:app_local_coherence_number_constraint} and \eqref{eq:app_neighbour_counting_constraint} reduce to the agreement condition in Eq.~\eqref{eq:local_agreement_filter_up_down}: the neighbouring occupation must agree in the unconjugated and conjugate fluid copies.

More generally, suppose that the paths $X_a(t)$ occupy distinct positions on a given time slice $t$. Equation \eqref{eq:app_local_coherence_number_constraint} then requires the conjugate coherences to occupy the same set of positions. Their matching to the unconjugated paths is specified by a permutation $\pi_t\in S_q$,
\begin{equation}
  \overline X_{\pi_t(a)}(t)=X_a(t).
  \label{eq:app_path_pairing}
\end{equation}
The pairing is therefore fixed between replica collisions.

We call a spacetime gate $j$ a \emph{replica collision} if it is visited by two or more paths $X_a(t)$. The coherences may occupy the same site or opposite sites of the bond. If a gate is visited by only one (unconjugated) trajectory $X_a(t)$, the local constraints above identify its conjugate partner uniquely on both sides of the gate. In that case $\pi_t$ is carried through the gate unchanged. Furthermore, Eq.~\eqref{eq:app_neighbour_counting_constraint} requires the neighbouring occupations in replica $a$ and conjugate replica $b=\pi_t(a)$ to agree. Thus, between replica collisions, each pair of unconjugated and conjugated fluid replicas obeys the same local conditioning as in the $q=1$ problem.

\subsubsection{Point-absorber reduction}

We next show that, at weak noise, the occupation constraint imposed by the phase average reduces to the same point-absorber rule in every replica, including at replica collisions. We first illustrate this for a collision of two coherence paths. Consider first a replica collision involving only two replicas $X_a$ and $X_b$.
Without loss of generality, label the replicas participating in the collision by $1,2,\overline 1,\overline 2$, where the barred labels denote the two conjugate replicas whose coherences visit the same spacetime gate.

Let $n_1,n_2,\overline n_{\overline 1},\overline n_{\overline 2} \in\{0,1\}$ denote the occupations neighbouring the four coherences. Eq.~\eqref{eq:app_neighbour_counting_constraint} imposes
\begin{equation}
  n_1+n_2
  =
  \overline n_{\overline 1}
  +
  \overline n_{\overline 2}.
  \label{eq:app_two_replica_occupation_filter}
\end{equation}
Equivalently, Eq.~\eqref{eq:app_neighbour_counting_constraint} imposes the projection
\begin{equation}
  \widehat{\mathcal Q}^{(2)}_j
  =
  \sum_{k=0}^{2}
  \Pi_k^{12}\,
  \overline\Pi_k^{\overline 1\overline 2},
  \label{eq:app_two_replica_phase_projector}
\end{equation}
where $\Pi_k^{12}$ projects onto configurations with exactly $k$ occupied neighbours among replicas $1$ and $2$, and $\overline\Pi_k^{\overline 1\overline 2}$ is the corresponding projector for the conjugate replicas. Note that the full gate action applies this projection and also allows the coherences to hop across the bond and for the pairing between the replicas $1,2$ and $\overline{1},\overline{2}$ to change. We can consider these effects separately. For now we focus on the effect of the projection onto compatible neighbouring occupations.

As in the main text, we introduce the overcomplete basis of diagonal operators,
\begin{equation}
  \dsym\equiv P^\downarrow,\qquad
  \usym\equiv P^\uparrow,\qquad
  \rsym\equiv(1-\nu)\dsym+\nu\usym.
  \label{eq:symbols}
\end{equation}
To see how this constraint acts on the bath-reset symbols $\rsym$, consider the local input in which all four neighbouring sites are in the state $\rsym=(1-\nu)\dsym+\nu\usym$. Applying Eq.~\eqref{eq:app_two_replica_phase_projector} gives
\begin{align}
  \widehat{\mathcal Q}^{(2)}_j
  \bigl(
    \rsym_1\rsym_2
    \rsym_{\overline 1}\rsym_{\overline 2}
  \bigr)
  ={}&
  (1-\nu)^4
  \dsym_1\dsym_2
  \dsym_{\overline 1}\dsym_{\overline 2}
  \nonumber\\
  &+
  \nu^2(1-\nu)^2
  \bigl(
    \usym_1\dsym_2+\dsym_1\usym_2
  \bigr)
  \bigl(
    \usym_{\overline 1}\dsym_{\overline 2}
    +
    \dsym_{\overline 1}\usym_{\overline 2}
  \bigr)
  \nonumber\\
  &+
  \nu^4
  \usym_1\usym_2
  \usym_{\overline 1}\usym_{\overline 2}.
  \label{eq:app_two_replica_reset_filter}
\end{align}
The three lines contain, respectively, zero, one, and two occupied neighbours.

As in the $q=1$ case, the leading weak-noise approximation neglects configurations containing $\usym$, i.e., retaining only the first line in Eq.~\eqref{eq:app_two_replica_reset_filter}. More generally, consider any input configuration $(s_1,s_2,s_{\overline 1},s_{\overline 2})$ with $s_j \in \{\dsym,\rsym\}$. Applying to this the projector in Eq.~\eqref{eq:app_two_replica_phase_projector} yields the output $(1-\nu)^{N_{\rsym}(j)} (\dsym_1\dsym_2 \dsym_{\overline 1}\dsym_{\overline 2})$ where $N_{\rsym}(j)$ is the number of $\rsym$ symbols in $(s_1,s_2,s_{\overline 1},s_{\overline 2})$.  This is precisely the result obtained by applying, independently in each replica participating in the collision, the absorption rule
\begin{equation}
  \rsym\longrightarrow(1-\nu)\dsym.
  \label{eq:app_q_replica_absorber}
\end{equation}
This naturally generalizes to collisions of $p>2$ coherence trajectories. Retaining only the occupation configurations (for the participating replicas at a collision) with occupation number $n=0$ yields closed dynamics on the symbols $\{\dsym,\rsym\}$ and is described by the point absorber rule Eq.~\eqref{eq:app_q_replica_absorber} at the coherence in each replica. Thus, to leading order at weak noise, the phase constraint acts independently in each participating replica through the same point absorber as in the $q=1$ problem.


\subsubsection{Validity of the point-absorber reduction}

It remains to verify that the omitted outputs containing $\usym$ symbols carry vanishing weight as $\gamma\to0$. We now reintroduce the neglected outputs, i.e., those producing $\usym$'s, and estimate the number of $\usym$ symbols in the stationary state. Within the point-absorber approximation, every fluid replica is locally described by the same quasi-stationary absorber problem as for $q=1$. The density $\varrho$ of $\rsym$ symbols vanishes at the coherence and varies across the void scale $\xi\sim\sqrt{D_0/\gamma}$ (Eq.~\eqref{eq:void_static_scalings_main}). The density at the coherence neighbour therefore scales as $\varrho_{\rm core}\sim\xi^{-1}\sim\sqrt{{\gamma}/{D_0}}.$

Consider first a gate visited by only one coherence trajectory $X_a$ (and its conjugate partner $\overline{X}_{\pi_t(a)}$). This scenario is captured exactly by Eq.~\eqref{eq:reset_filter_rules_main} in the main text. The only outputs with a $\usym$ present necessarily have one in both the unconjugated and conjugated replicas, (denoted $\usym \boxtimes \usym$ in the main text). To produce this output requires an $\rsym$ to be present in both the unconjugated and conjugated replicas at the coherence. Such pairs are therefore produced at a rate $\Gamma_{\usym} \sim \varrho_{\rm core}^{\,2} \sim{\gamma}/{D_0}$ (up to an order-one factor depending on $\nu$).

Once produced, the two $\usym$ symbols diffuse away from the coherence. If either one returns without the other returning to the same gate at the same time, the neighbouring occupations fail to agree with probability approaching unity as $\gamma \to 0$. Indeed, the fluid in the other replica is in the state $\dsym$ next to the coherence with probability $1-O(\varrho_{\rm core})$. The history is then removed by the phase constraints. To leading order, the long-lived histories containing such a pair are therefore those in which neither $\usym$ has returned to the coherence. For a one-dimensional diffusing symbol initially near an absorbing point, the no-return probability up to time $t$ scales as $t^{-1/2}$. The survival probability of the pair consequently scales as $P_{\usym}^{\rm surv}(t) \sim t^{-1}$. 

A bath reset removes a $\usym$ on the timescale $\gamma^{-1}$. The effective residence time of a produced pair in the quasi-stationary ensemble is therefore
\begin{equation}
  \tau_{\usym}^{\rm eff}
  \sim
  \int_1^{\gamma^{-1}}dt\,
  P_{\usym}^{\rm surv}(t)
  \sim
  \log(1/\gamma).
  \label{eq:app_u_effective_residence_time}
\end{equation}
Combining the production rate and residence time gives the expected number of $\usym$'s in the stationary state
\begin{equation}
  N_{\usym}^{\rm QSS}
  \sim
  \Gamma_{\usym}\tau_{\usym}^{\rm eff}
  \sim
  \frac{\gamma}{D_0}\log(1/\gamma)
  \ll1.
  \label{eq:app_u_number_no_collision}
\end{equation}
This repeats the $q=1$ estimate in the End Matter (see also Ref.~\cite{previous}).

The exceptional processes omitted from this estimate do not modify the leading long-time survival probability or $N_{\usym}^{{\rm QSS}}$. Firstly, we consider the exceptional case where both $\usym$'s return to the coherence at the same time. In this case, the equal occupation condition imposed by the phase average is satisfied. Conditional on a return at time $t$, however, the probability that the other diffusing symbol is also at the coherence is $O(t^{-1/2})$. Almost every return therefore occurs alone and removes the history. Furthermore, after a simultaneous return, both $\usym$ symbols remain present and their subsequent returns are again subject to the survival probability cost. Histories containing such simultaneous returns consequently give only a relative $O(t^{-1/2})$ correction to the leading result, $P_{\usym}^{\rm surv}(t)\sim t^{-1}$. The second exceptional case is when one $\usym$ returns and is accompanied by an $\rsym$ in another replica. This occurs with probability $O(\varrho_{\rm core})=O(\sqrt{\gamma})$ (conditional on the return of a $\usym$). Such an event then produces an additional $\usym$ symbol, with an additional no-return probability cost. This case is then likewise suppressed by a further factor of $1/\sqrt{t}$ (before the bath reset time $\gamma^{-1}$).

We next check that replica collisions do not alter this scaling. Consider a collision involving $p$ unconjugated coherence paths $\{ X_a \}$ and the corresponding $p$ conjugate paths $\{ \overline{X}_{\pi(a)}\}$. An omitted output containing a $\usym$ in one unconjugated replica $a$ and one $\usym$ in a conjugate replica $\overline b$ requires an $\rsym$ to be present in both of these fluid replicas. For each possible pair $(a,\overline b)$, the production probability is therefore again $O(\varrho_{\rm core}^{\,2})=O(\gamma/D_0)$. There are at most $\Ord(p^2)$ such choices, so for any fixed $p\leq q$ a replica collision changes only the prefactor of the production rate, not its dependence on $\gamma$.

Now suppose that the $\usym$ in replica $a$ subsequently returns to its coherence at a gate involved in a collision of $p$ replicas. Unless the other $\usym$ (of the produced pair) returns to the same gate simultaneously, the history can survive only if at least one of the conjugate fluid replicas participating in the collision contains an $\rsym$ symbol. The probability of finding an $\rsym$ in at least one of these replicas is $1-(1-\varrho_{\rm core})^p =O\!\left(p\varrho_{\rm core}\right) \sim p\sqrt{\gamma}$. This has the same $\gamma$ scaling for all $p$, and in particular, the same as in the $q=1$ problem. Thus, at weak noise, the point absorber approximation in Eq.~\eqref{eq:app_q_replica_absorber} remains valid for general $q$.

\subsubsection{Feynman-polaron reduction of the balanced moments at weak noise}

The discussion above is independent of the circuit geometry and applies equally to brickwork and Poissonized circuits. For convenience, we now specialize to a Poissonized circuit, in which two-site gates act at independent random times and replica collisions occur at isolated gate events.

The replicated statistical-mechanics evolution does not generally define a nonnegative path measure, so the Feynman--Jensen variational argument~\cite{FeynmanPolarCrystal} cannot be applied directly to the complete replicated history. Between collisions, however, the pairing is fixed and the fluid weight reduces to a product of positive $q=1$ problems. We therefore apply the Feynman-polaron replacement separately on each collision-free segment, before joining the segments and resumming the collision histories.

Within the point-absorber approximation, the coherence amplitude no longer depends on the fluid history, since the site neighbouring the coherence is always retained in the state $\dsym$. We therefore define
\begin{equation}
  \mathcal A_U^{(0)}[X]
  =
  \prod_{j\in X}c_{j,a_j}(0),
  \label{eq:app_vacuum_coherence_amplitude}
\end{equation}
where $a_j\in\{{\rm s},{\rm h}\}$ denotes the stay or hop prescribed by the path $X$. Summing independently over the fluid histories in each replica gives
\begin{align}
  M_{q,U} \simeq
  \sum_{\{X_a\}}
  \sum_{\{\overline X_{a}\}}
  \left[
    \prod_{a=1}^q
    Z_U[X_a]
    \prod_{b=1}^q
    Z_U[\overline X_{b}]
  \right]\times
  \mathbb E_\phi\!\left[
    \prod_{a=1}^q
    \mathcal A_U^{(0)}[X_a]
    \prod_{b=1}^q
    \mathcal A_U^{(0)}
      [\overline X_{b}]^*
  \right].
  \label{eq:app_phase_averaged_factorized_model}
\end{align}
The fluid weights have therefore factorized completely. 

The main text already determines the leading random free energy fluctuations generated by the quenched random hopping probabilities. To isolate the remaining phase-disorder contribution, we treat the hopping probabilities as annealed, characterized by a uniform diffusivity $D_0$. The only remaining disorder is therefore the gate phases $\{\phi_j\}$. We henceforth write the hopping-annealed balanced moment, fluid partition sum, and coherence path amplitude,
\begin{equation}
    M_{q,U} \to M_q, \qquad Z_U \to Z, \qquad \mathcal A_{U} \to \mathcal A_{\phi}.
    \label{eq:hopping-annealed}
\end{equation}
We restore the effects of hopping disorder at the end of the appendix.

In the Poissonized circuit, replica collisions occur at isolated gate times. For a fixed collision history
\begin{equation}
  h=
  \left(
    t_1,\ldots,t_k;
    \pi_0,\ldots,\pi_k
  \right),
\end{equation}
the pairing is fixed to $\pi_i\in S_q$ on each open interval $I_i=(t_i,t_{i+1})$. The phase average locks
\begin{equation}
  \overline X_{\pi_i(a)}(\tau)=X_a(\tau),
  \qquad
  \tau\in I_i,
  \label{eq:app_poisson_segment_locking}
\end{equation}
while the instantaneous gate at $t_i$ supplies the exact coherence matrix element
\begin{equation}
  \left[\widehat{\mathcal T}^{(0)}_{t_i}\right]_{(\{X(t_i^+)\},\pi_i), (\{X(t_i^-)\},\pi_{i-1})},
\end{equation}
where $\widehat T_{\tau_i}^{(0)}$ is the exact $q$-replica transfer matrix for the gate at $\tau_i$ in the $n=0$ occupation sector, supplying the amplitudes for the transition $(\{X(t_i^-)\},\pi_{i-1})\to (\{X(t_i^+)\},\pi_i)$. Here, $\{X(t_i^-)\}$ ($\{X(t_i^+)\}$) is the configuration of coherence positions immediately before (after) the collision.

The phase-averaged coherence weight can therefore be resolved as
\begin{align}
  \mathbb E_\phi\!\left[
    \prod_a\mathcal A_\phi^{(0)}[X_a]
    \mathcal A_\phi^{(0)}
      [\overline X_{a}]^*
  \right]=
  \sum_h
  \Delta_h[\{\overline X\}|\{X\}]
  \left[
    \prod_{i=0}^{k}
    \prod_{a=1}^q
    P_{I_i}^{(0)}[X_a]
  \right]
  \prod_{i=1}^{k}
  \left[\widehat{\mathcal T}^{(0)}_{t_i}\right]_{(\{X(t_i^+)\},\pi_i), (\{X(t_i^-)\},\pi_{i-1})},
  \label{eq:app_poisson_collision_resolution}
\end{align}
where $P_{I_i}^{(0)}[X_a]=|\mathcal A_{\phi,I_i}^{(0)}[X_a]|^2$ is the probability for the coherence path $X_a$ on interval $I_i$, and $\Delta_h[\{\overline X\}|\{X\}]$ is the indicator function for the collision and pairing history $h$:
\begin{equation}
  \Delta_h[\{\overline X\}|\{X\}]
  =
  \begin{cases}
    1,
    &
    \begin{array}{l}
      \{X_a\}\text{ has collisions precisely at }
      t_1,\ldots,t_k,\\[-2pt]
      \overline X_{\pi_i(a)}(\tau)=X_a(\tau)
      \text{ for all }\tau\in I_i
    \end{array}
    ,
    \\[10pt]
    0,
    & \text{otherwise}.
  \end{cases}
  \label{eq:app_poisson_history_locking}
\end{equation}
The indicators form a resolution of unity,
$\sum_h\Delta_h[\{X\}|\{\bar X\}]=1$.

We now insert complete sets of fluid configurations at the collision times. On the support of $\Delta_h$, the paired fluid and coherence weight on each interval reduces to the $q=1$ problem, enabling us to make the same Feynman-polaron approximation we made for $q=1$ in the main text (see also Ref~\cite{previous}). We do this now, while keeping the coherence path probabilities $P_{I_i}^{(0)}[X_a]$ explicit:
\begin{equation}
  P_{I_i}^{(0)}[X_a]\,
  Z_{I_i}[X_a]\,
  Z_{I_i}[\overline X_{\pi_i(a)}] = P_{I}^{(0)}[X]Z_{I}[X]^2
  \simeq
  P_{I}^{(0)}[X]
  \int\mathcal D y\,\mathcal D\overline y\,
  e^{-S_{\mathrm F,I}[X,y]
     -S_{\mathrm F, I}[X,\overline y]}.
  \label{eq:app_segment_q1_feynman_replacement}
\end{equation}
Here $S_{\mathrm F}$ denotes the single-copy fluid contribution obtained from the $q=1$ variational calculation~\eqref{eq:feynman_eff_action_end},
\begin{equation}
    S_F[X,y] = \int d\tau \left[\lambda_0 + \frac{\dot{y}^2}{8D_{\rm void}}+\frac{D_0}{8\ell^4} z^2\right].
\end{equation}
Here the coherence propagators can be retained exactly: the purpose of the Feynman-polaron ansatz~\cite{FeynmanPolarCrystal} is to make a variational estimate for the fluid action, for which all we require is that $P_{\phi,I}^{(0)}[X]$ is diffusive with diffusivity $D_0$.

Applying Eq.~\eqref{eq:app_segment_q1_feynman_replacement} on every interval of a fixed collision history $h$ gives
\begin{align}
  &\Delta_h[\{\overline X\}|\{X\}]
  \left[
    \prod_{a=1}^q \left(\prod_{i=0}^{k}
    P_{I_i}^{(0)}[X_a]\right) Z[X_a] Z[\overline X_{a}]
  \right]
  \nonumber\\
  &\quad\simeq
  \Delta_h[\{\overline X\}|\{X\}]
  \left[
    \prod_{i=0}^{k}\prod_{a=1}^q
    P_{I_i}^{(0)}[X_a]
  \right]
  \int
  \prod_{a=1}^q \mathcal D y_a
  \mathcal D\overline y_{a}
  \exp\left[
    -\sum_{i=0}^{k}\sum_{a=1}^q
      \left(S_{\mathrm F, I_i}[X_a,y_a]+S_{\mathrm F, I_i}
      [X_a,\overline y_{\pi_i(a)}]\right)
  \right].
  \label{eq:app_fixed_history_segment_replacement}
\end{align}
Note that this decomposition into collision-free intervals does not assume that the void re-equilibrates following each collision. The collective coordinates are propagated continuously through the collision times by integrating over their common intermediate endpoints (i.e., $\int dy\, |y\rangle\langle y|=\mathds 1$). The decomposition serves only to resolve the changing replica pairing. 



On the support of $\Delta_h$, i.e., for a fixed collision history, we may reorganize the terms in the exponential of Eq.~\eqref{eq:app_fixed_history_segment_replacement},
\begin{align}
  \sum_{i,a}
  S_{\mathrm F, I_i}
  [X_a,\overline y_{\pi_i(a)}]
  =
  \sum_{i,a}
  S_{\mathrm F,I_i}
  [\overline X_{\pi_i(a)},
   \overline y_{\pi_i(a)}]=
  \sum_{\overline a=1}^q
  S_{\mathrm F}
  [\overline X_{\overline a},
   \overline y_{\overline a}],
   \qquad \sum_{i,a}
  S_{\mathrm F,I_i}[X_a,y_a]
  =
  \sum_{a=1}^q
  S_{\mathrm F}[X_a,y_a].
  \label{eq:app_compose_conjugate_feynman_action}
\end{align}

Thus, the exponential in Equation~\eqref{eq:app_fixed_history_segment_replacement} is independent of the collision history. Substituting the collision resolution Eq.~\eqref{eq:app_poisson_collision_resolution} into Eq.~\eqref{eq:app_phase_averaged_factorized_model}, and using Eq.~\eqref{eq:app_compose_conjugate_feynman_action}, now gives
\begin{align}
  M_q
  \simeq
  \sum_{\{X_a\}}
  \sum_{\{\overline X_{a}\}}
  \int
  \prod_a\mathcal D y_a
  \mathcal D\overline y_{a}
  \exp\left[
    -\sum_a S_{\mathrm F}[X_a,y_a]
    -\sum_a S_{\mathrm F}[\overline X_{a},\overline y_{a}]
  \right]\times
  \mathbb E_\phi\!\left[
    \prod_a\mathcal A_\phi^{(0)}[X_a]
    \mathcal A_\phi^{(0)}
      [\overline X_{a}]^*
  \right]
  .
  \label{eq:app_explicit_recombined_history_sum}
\end{align}

Thus, the coherence trajectories and their collision weights can be retained exactly. Only the fluid dynamics attached to each trajectory has been replaced by the effective Feynman-polaron description obtained for $q=1$.

\subsection{Mapping to directed waves}

The preceding reduction expresses all balanced moments in terms of a single effective phase-disordered amplitude. We now separate its slow void coordinate from the confined internal motion, determine the coarse-grained effect of the random phases, and finally restore the hopping disorder.

\subsubsection{Single-replica path sum}

Equation~\eqref{eq:app_explicit_recombined_history_sum} is the balanced moment of a single-replica amplitude,
\begin{equation}
  M_{q}
  \simeq
  \mathbb E_\phi
  \left[
    \left|G_\phi^{\mathrm F}\right|^{2q}
  \right],
  \qquad
  G_\phi^{\mathrm F}
  =
  \sum_X
  \int\mathcal D y\,
  e^{-S_{\mathrm F}[X,y]}
  \mathcal A_\phi^{(0)}[X].
  \label{eq:app_feynman_balanced_moment}
\end{equation}
Writing $X=y+z$ and taking the continuum limit gives
\begin{align}
  G_\phi^{\mathrm F}
  =
  \int\mathcal D y\,\mathcal D z\,
  \exp\Bigg\{
    -\int d\tau
    \left[
      \frac{\dot y^2}{8D_{\mathrm{void}}}
      +
      \frac{D_0}{8\ell^4}z^2
      +
      \lambda_0
    \right]\Bigg\}
    \mathcal A_\phi^{(0)}[X].
  \label{eq:app_feynman_effective_amplitude}
\end{align}

For uniform hopping probabilities appropriate for annealed hopping disorder (Eq.~\eqref{eq:hopping-annealed}), we have
\begin{equation}
  \mathcal A_\phi^{(0)}[X]
  =
  \sqrt{P_0[X]}\,e^{i\Phi_\phi[X]},
  \qquad
  \sqrt{P_0[X]}
  \simeq
  \exp\left[
    -\int d\tau\,\frac{\dot X^2}{8D_0}
  \right],
  \qquad
  \Phi_\phi[X]
  =
  \int d\tau\,\phi(X(\tau),\tau).
  \label{eq:app_bare_phase_amplitude}
\end{equation}

The internal coordinate relaxes on the timescale
$\tau_\ell\sim\ell^2/D_0$. During this time,
\begin{equation}
  \frac{\sqrt{D_{\rm void}\tau_\ell}}{\ell}
  \sim
  \sqrt{\frac{D_{\rm void}}{D_0}}
  \ll1,
\end{equation}
so that $y$ may be treated as fixed while integrating out $z$. The corresponding internal path sum is
\begin{equation}
  \mathcal W_\phi[y]
  =
  \int\mathcal D z\,
  \exp\left\{
    -\int d\tau
    \left[
      \frac{\dot z^2}{8D_0}
      +
      V_\ell(z)
    \right]
    +
    i\int d\tau\,
    \phi(y(\tau)+z(\tau),\tau)
  \right\},
  \qquad
  V_\ell(z)=\frac{D_0}{8\ell^4}z^2.
  \label{eq:app_internal_phase_path_sum}
\end{equation}
The $\dot y$-dependent terms generated by $\dot X=\dot y+\dot z$ give only subleading corrections to the slow kinetic action for $y$ and will be neglected.

\subsubsection{Phase-induced damping fluctuations}

We now determine the effects of the microscopic phase disorder on the slow polaron coordinate $y$ upon integrating out the confined internal motion.

Consider a spacetime block $B$ of spatial width $\Ord(\ell)$ and duration $T$ satisfying
\begin{equation}
  \tau_\ell
  \ll
  T
  \ll
  \frac{\ell^2}{D_{\rm void}}.
  \label{eq:app_internal_block_window}
\end{equation}
Within this window the internal motion has reached its stationary regime, while the void center remains effectively fixed. We write its block weight as
\begin{equation}
  \mathcal W_{\phi,B}(y)
  =
  \exp\left[
    -\lambda_{\rm int}T
    -\delta F_{\phi,B}(y)
    +i\theta_{\phi,B}(y)
  \right].
  \label{eq:app_internal_block_weight}
\end{equation}

The block phase is uniformly distributed. Indeed, a common shift of the microscopic phases within the block, $\phi(x,\tau)\rightarrow \phi(x,\tau)+\delta\Theta/T$, multiplies every internal trajectory by $e^{i\delta\Theta}$. Invariance of the gate ensemble therefore implies $\theta_{\phi,B}(y)\sim{\rm Unif}[0,2\pi)$. Because the microscopic phase disorder is short ranged and the confined particle transfer operator has a gap $\Ord(D_0/\ell^2)$, residual correlations decay beyond $\ell$ in space and $\tau_\ell$ in time.

To quantify the realization-dependent damping produced by the random phases, we consider the fluctuations of the norm of the trapped internal wavefunction. Fixing $y=0$, let $\psi_\phi(z,T)$ denote the wavefunction obtained by evolving an initial internal state for time $T$ under the non-Hermitian trapped random-phase dynamics defined by Eq.~\eqref{eq:app_internal_phase_path_sum}, and define
\begin{equation}
  N_\phi(T)
  =
  \sum_z
  \left|\psi_\phi(z,T)\right|^2.
\end{equation}

The first and second moments of $N_\phi$ are generated by the phase-averaged one- and two-replica transfer operators, $\mathsf T_1$ and $\mathsf T_2$, respectively. For uniform hopping probabilities, which we assume until hopping disorder is restored below, we write $\mathsf T_2=\mathsf T_1\otimes\mathsf T_1+\mathsf C$, where $\mathsf C$ is a local interaction acting when the two replicas ($z_1$ and $z_2$) encounter the same gate. For $T\gg\tau_\ell$,
\begin{equation}
  \mathbb E_\phi[N_\phi(T)]
  \asymp
  e^{-\Lambda_1T},
  \qquad
  \mathbb E_\phi[N_\phi(T)^2]
  \asymp
  e^{-\Lambda_2T},
  \label{eq:app_internal_norm_moments}
\end{equation}
so that the growth rate of the relative norm fluctuations is
\begin{equation}
  \kappa_\ell
  \equiv
  2\Lambda_1-\Lambda_2.
  \label{eq:app_internal_kappa_definition}
\end{equation}

The long-wavelength structure of $\mathsf C$ is most easily obtained from the underlying discrete quantum walk, before adding the confining potential and taking the continuum limit. Let $I_x=|\psi_x|^2$. In this unconfined problem, unitarity gives $\sum_x I_x=1$ in every realization. Define also the intensity-intensity correlation function $P_{x_1x_2}=\mathbb E_\phi[I_{x_1}I_{x_2}]$. Since $\sum_x I_x=1$, we also have $\sum_{x_2}P_{x_1x_2}=\mathbb E_\phi\big[I_{x_1}\sum_{x_2}I_{x_2}\big]=\mathbb E_\phi[I_{x_1}]$. Marginalizing either replica of the two-replica evolution must therefore reproduce the one-replica evolution. Denoting the uniform state by $\langle\mathbf 1|=\sum_x\langle x|$, this implies
\begin{equation}
  (\mathbb I\otimes\langle\mathbf 1|)\mathsf T_2
  =
  \mathsf T_1(\mathbb I\otimes\langle\mathbf 1|),
\end{equation}
together with the corresponding identity for the other replica. Since $\mathsf T_1\otimes\mathsf T_1$ obeys the same marginal identities,
\begin{equation}
  (\mathbb I\otimes\langle\mathbf 1|)\mathsf C
  =
  (\langle\mathbf 1|\otimes\mathbb I)\mathsf C
  =
  0.
\end{equation}
Furthermore, the transfer operators are symmetric, so the analogous contractions on the incoming side also vanish. Thus $\mathsf C$ annihilates a wavefunction that is uniform in any incoming or outgoing replica coordinate. Locality then implies that its leading long-wavelength matrix element contains one gradient on each of its four legs,
\begin{equation}
  \langle k_1',k_2'|\mathsf C|k_1,k_2\rangle
  \propto
  D_0\,
  k_1'k_2'k_1k_2
  +\cdots,
  \label{eq:app_collision_gradient_structure}
\end{equation}
with momentum conservation understood.

Let $\varphi_0(z)=\ell^{-1/2}f(z/\ell)$ be the normalized one-replica eigenfunction in the trap. First-order perturbation theory about the product state $\varphi_0\otimes\varphi_0$ then gives
\begin{equation}
  \kappa_\ell
  \sim
  D_0
  \sum_z
  \left[\Delta\varphi_0(z)\right]^4
  \sim
  {D_0}/{\ell^5},
  \label{eq:app_internal_kappa_scaling}
\end{equation}
up to an $\Ord(1)$ coefficient, where $\Delta$ is a microscopic lattice difference. Since the trapped excitation gap is $\Ord(D_0/\ell^2)$, the ratio of this shift to the gap is $\Ord(\ell^{-3})\ll1$, controlling the perturbative calculation.

The relative variance is consequently
\begin{equation}
  \mathbb E_\phi\!\left[\delta_\phi(T)^2\right]
  =
  \frac{
    {\rm Var}_\phi[N_\phi(T)]
  }{
    \mathbb E_\phi[N_\phi(T)]^2
  }
  \asymp
  e^{\kappa_\ell T}-1
  \simeq
  \kappa_\ell T,
  \quad {\rm where} \ \
  \delta_\phi(T)
  \equiv
  \frac{N_\phi(T)}{\mathbb E_\phi[N_\phi(T)]}-1.
  \label{eq:app_internal_relative_variance}
\end{equation}
Here we used $\kappa_\ell T\ll1$ throughout the window~\eqref{eq:app_internal_block_window}. Thus $\delta_\phi=O_{\rm rms}(\sqrt{\kappa_\ell T})\ll1$.

The fluctuating part of the single-amplitude damping, $F_\phi^{\rm int}(T)\equiv -\frac12\log N_\phi(T),$ is therefore
\begin{equation}
  F_\phi^{\rm int}(T)
  -
  \mathbb E_\phi[F_\phi^{\rm int}(T)]
  =
  -\frac12
  \left[
    \log(1+\delta_\phi)
    -
    \mathbb E_\phi\log(1+\delta_\phi)
  \right].
\end{equation}
Expanding $\log(1+\delta_\phi)=\delta_\phi-\delta_\phi^2/2+\cdots$ and using $\mathbb E_\phi[\delta_\phi]=0$ gives, to leading order,
\begin{equation}
  {\rm Var}_\phi[F_\phi^{\rm int}(T)]
  =
  \frac14\,
  \mathbb E_\phi[\delta_\phi(T)^2]
  +o(\kappa_\ell T)
  \sim
  \kappa_\ell T.
  \label{eq:app_internal_log_norm_variance}
\end{equation}

Blocks separated by more than the internal correlation scales $\ell$ and $\tau_\ell$ are only weakly correlated. Summing the fluctuating free energy over such blocks up to the slow coarse-graining time $\widetilde\tau \sim{\xi^2}/{D_{\rm void}}\sim{\ell^{9/2}}/{D_0}$, therefore gives the central-limit estimate
\begin{equation}
  {\rm Var}_\phi(\delta F_{\phi})
  \sim
  \kappa_\ell\widetilde\tau
  \sim
  \ell^{-1/2}\implies {\rm std}_{\phi}[\delta F_{\phi}]
  \sim
  \ell^{-1/4}
  \sim
  \left(\frac{\gamma}{D_0}\right)^{1/12}
  \ll1.
  \label{eq:app_internal_phase_damping_fluctuation}
\end{equation}

The phase-disorder-induced free-energy fluctuation is therefore parametrically smaller than the $\Ord(1)$ contribution from transport disorder on the same coarse-grained spacetime block. 

\subsubsection{Restoring hopping disorder}

We have found that phase disorder produces an order-one random block phase but only a parametrically small fluctuation of the real damping rate. We now restore the quenched hopping disorder, whose fluctuations we have already seen (in the main text) produce an $\Ord(1)$ real random energy on the $(\xi,\widetilde\tau)$ scale. We now combine the coarse-grained phase disorder and random energy fluctuations into a complex random potential.

As explained in the main text, the leading effect of restoring the hopping disorder,  $D_U(x,\tau)=D_0+\delta D_U(x,\tau)$, is to promote the quasi-stationary decay rate $\lambda_0$ to a fluctuating decay rate $\lambda(y,\tau) = \lambda_0 + V_\xi(y,\tau)/2$ (Eq.~\eqref{eq:random-potential}). Over a spacetime block of width $\xi$ and duration $\widetilde\tau$, this produces an $\Ord(1)$ real free-energy fluctuation, whereas Eq.~\eqref{eq:app_internal_phase_damping_fluctuation} shows that the real fluctuation generated by the internal phase disorder is parametrically smaller. The latter may therefore be neglected, while its uniformly distributed block phase must be retained.

Introducing the coarse-grained coordinates
\begin{equation}
  Y={y}/{\xi},
  \qquad
  s={2\tau}/{\widetilde\tau},
\end{equation}
and denoting the deterministic decay rate by $\overline\lambda=\lambda_0+\lambda_{\rm int}$, the resulting long-wavelength amplitude is
\begin{align}
  G_U^{\mathrm F}
  \simeq
  e^{-\overline\lambda t}
  \int\mathcal D Y\,
  \exp\left\{
    -\int ds
    \left[
      \frac{(\partial_sY)^2}{4}
      +
      \mathcal V_U(Y(s),s)
    \right]
  \right\},
  \label{eq:app_directed_wave_path_sum}
\end{align}
Here $\mathcal V_U$ is a short-range correlated complex random potential. Its real part is dominated at weak noise by the transport-disorder contribution $V_\xi/2$, while its imaginary part represents the coarse-grained random phases. On scales large compared with one block its correlations may be represented as
\begin{equation}
  \mathbb E\left[
    \mathcal V_{\alpha}(Y,s)
    \mathcal V_{\beta}(Y',s')
  \right]_{\rm c}
  =
  \Delta_{\alpha\beta}
  \delta(Y-Y')\delta(s-s'),
  \qquad
  \alpha,\beta\in\{\mathrm R,\mathrm I\},
\end{equation}
where $\mathcal V_R = {\rm Re}[\mathcal V]$ and $\mathcal V_I = {\rm Im}[\mathcal V]$, and where $\Delta_{\alpha\beta}=\Ord(1)$.

Equivalently, after removing the deterministic decay, Eq.~\eqref{eq:app_directed_wave_path_sum} is the Feynman--Kac representation of the equation for a directed wave in a random medium~\cite{PhysRevLett.62.941,PhysRevLett.66.2176,PhysRevA.45.8859,BLUM1992588,1994cond.mat.11022K,PhysRevB.83.052405},
\begin{equation}
  \partial_s G_U^{\mathrm F}
  =
  \partial_Y^2 G_U^{\mathrm F}
  -
  \mathcal V_U(Y,s) G_U^{\mathrm F}.
  \label{eq:app_directed_wave_equation}
\end{equation}

Therefore, within the projected-dynamics ansatz, $G_U^F$ matches all moments~\footnote{Since unbalanced moments vanish both for $G_U^F$ and $G_U$.} of the microscopic Green function upon coarse-graining, up to corrections that vanish in the scaling limit. For every fixed replica number $q$, at fixed rescaled coordinates and in the weak-noise limit,
\begin{equation}
  \mathbb E_U\!\left[
    |G_U|^{2q}
  \right]_{\rm coarse-graining}
  \simeq
  \mathbb E_{\mathcal V}\!\left[
    |G_U^{\rm F}|^{2q}
  \right].
  \label{eq:app_balanced_moment_matching}
\end{equation}
Here the average on the right is over the effective complex random potential $\mathcal V_U$.

Balanced moments were introduced precisely to encode the sample-to-sample statistics of the complex Green function. Moreover, the unbalanced moments vanish because of the uniformly distributed phase. Equation~\eqref{eq:app_balanced_moment_matching} therefore shows that, within the projected-dynamics ansatz, $G_U^{\rm F}$ reproduces the coarse-grained statistics of typical realizations of $G_U$, rather than only an annealed observable such as $\mathbb E_U[|G_U|^2]$. In this sense, $G_U^{\rm F}$ describes a typical single-particle Green function under coarse-graining.

\end{document}